\documentclass{IEEEoj}
\usepackage{cite}
\usepackage{amsmath,amssymb,amsfonts}
\usepackage{algorithmic}
\usepackage{graphicx,color}
\usepackage{textcomp}
\usepackage{mathabx}
\usepackage{enumerate}
\usepackage[]{footmisc}
\usepackage{cite}
\usepackage{resizegather}
\newcounter{daggerfootnote}
\usepackage{mathtools}

\usepackage{tikz}
\usepackage{amsthm}
\newtheorem{theorem}{Theorem}

\newtheorem{lemma}[theorem]{Lemma}

\usepackage[hidelinks]{hyperref}
\usepackage{booktabs}
\def\BibTeX{{\rm B\kern-.05em{\sc i\kern-.025em b}\kern-.08em
    T\kern-.1667em\lower.7ex\hbox{E}\kern-.125emX}}
\renewcommand{\appendixname}{APPENDIX}

\begin{document}

\doiinfo{10.1109/OJCOMS.2022.3219557}
\DeclareRobustCommand*{\IEEEauthorrefmark}[1]{%
  \raisebox{0pt}[0pt][0pt]{\textsuperscript{\footnotesize\ensuremath{#1}}}}
\title{Interference-Aware Accurate Signal Recovery in sub-1 GHz UHF Band Reuse-1 Cellular OFDMA Downlinks}
\author{ABHAY MOHAN M. V.\textsuperscript{\href{https://orcid.org/0000-0002-9290-9929}{\includegraphics[scale=.04]{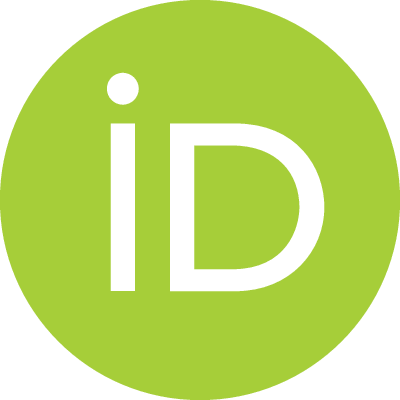}}} \authorrefmark{1}, and K. GIRIDHAR\textsuperscript{\href{https://orcid.org/0000-0001-6044-2036}{\includegraphics[scale=.04]{ORCIDiD_iconvector.eps}}}\authorrefmark{1}}
\affil{Department of Electrical Engineering, Indian Institute of Technology Madras, Chennai 600 036, India}
\corresp{CORRESPONDING AUTHOR: Abhay Mohan M. V. (e-mail: abhay@telwise-research.com).}
\markboth{Interference-Aware Accurate Signal Recovery in sub-1 GHz UHF Band Reuse-1 Cellular OFDMA Downlinks}{Abhay Mohan M. V. and K. Giridhar}

\begin{abstract}
Reuse-1 systems operating in the sub-1 GHz UHF band are limited by substantial co-channel interference (CCI). In such orthogonal frequency division multiple access (OFDMA) cellular systems, the inter-sector or inter-tower interference (ITI) makes accurate signal recovery quite challenging as sub-1 GHz bands only support single-input single-output (SISO) links. Interference-aware receiver algorithms are essential to mitigate the ITI in such low-frequency bands. Such algorithms enable ubiquitous mobile broadband access over the entire homeland, say with $>95\%$ geographical coverage with quality of service guarantees. One element of the interference-aware signal recovery is the least-squares-based joint channel estimation scheme that uses non-orthogonal pilot subcarriers. This estimator is then compared with a variant that uses orthogonal pilot subcarriers to bring out the advantage of this joint estimator. It is shown that the proposed joint estimator requires fewer pilots to be well-determined when compared to its under-determined orthogonal counterpart. Moreover, it is easy to implement and does not require any knowledge of channel statistics. This work also derives a compensation factor needed for the interference-aware detector in the presence of inter-carrier interference (ICI) originating from multiple transmitters. Simulation results show that the proposed joint channel estimator outperforms traditional estimators at moderate to high frequency selectivity. The proposed compensation factor to the joint detector is found to be essential for recovering the transmitted signal in the absence of phase-tracking pilots.
\end{abstract}


\begin{IEEEkeywords}
Carrier frequency offset compensation, inter-carrier interference, interference-aware receivers, inter-tower interference, joint channel estimation,  joint detection, OFDMA, UHF cellular systems.
\end{IEEEkeywords}


\maketitle

\section{INTRODUCTION}
\IEEEPARstart{B}{roadband} cellular networks based on the current 4G-Long Term Evolution (LTE) or the emerging 5G-New Radio (NR) wireless standards use universal frequency reuse or reuse-1 \cite{ameen2020proposed,qi2021energy}.  In such OFDM/OFDMA block-modulated networks, all cell towers and sectors use the same frequency resource\footnote{The word sector is used to refer to the full $360^{\circ}$ region served by a cell site, or to simply one part of that region (typically a $120^{\circ}$ portion, when three sectors are deployed per cell site). In this paper, the term inter-tower interference (ITI) also subsumes the inter-sector interference that could be present at the sector boundaries.} to provide a higher sum throughput. Co-channel interference (CCI), which arises in such ultra-dense networks, is a crucial bottleneck that restricts the throughput of the cell-edge user \cite{cao2018interference}. CCI manifests as inter-cell or inter-sector interference, broadly referred to as inter-tower interference (ITI). This contribution is focussed on a downlink cellular model with ITI. Downlink power control, interference coordination, beamforming, etc. (see, for example, \cite{Xiao2020Reinforcement, Wang2020Decentralized, Antwi2019Evaluation, necker2006towards,  Moon2019Hybrid}) are used to reduce this ITI and improve cell-edge user throughput.

OFDMA networks deployed in the ultra-high frequency (UHF), particularly the sub-1 GHz bands, can provide excellent geographical coverage and are used for supplementary uplink in 5G \cite{3gpp2020overall}, \cite{3gpp381011}. The 3GPP study item \cite{3gpp38913} has identified typical deployment scenarios for next-generation access technologies. Eleven of the twelve scenarios identified have an option for sub-6 GHz deployment, and three particularly consider sub-1 GHz bands alone. The scenarios include rural deployment with inter-site distances or ISD of 1 to 5 km, extreme long-distance coverage in low density areas with ISD of the order of 100 km, and urban coverage for massive machine-type communications (mMTC). The importance of these bands for rural connectivity is also studied in \cite{Koratagere2021Techno} and \cite{Jeon2021MIMO}. This frequency band has also been examined for suitability for wideband Short-Range Devices (SRDs) in \cite{etsi103245},\cite{ECC189} as well as low-power, long-range IoT \cite{Hussain2021ACompact}.  The band also finds use in deploying resilient machine-to-machine systems, including smart grids, which are critical to the national infrastructure \cite{etsi103492}. The IEEE 802.22 and 802.11af standards, which are expected to play a significant role in bridging the rural connectivity for 6G networks, also operate in this band \cite{Yaacoub2020AKey}, \cite{EriccBlog}.
 
 However, the sub-1 GHz bands do not support spatial filtering/beamforming because the large wavelengths involved make single input single output (SISO) the only realistic deployment. The receiver will likely pick up significant ITI, especially in the cell-edge region, even after employing interference coordination and power control algorithms. The ITI is especially detrimental if the aim is to provide coverage $>95\%$. In such cases, cells have to be designed in an overlapping manner. This overlap causes high CCI at the cell boundaries and has to be overcome purely by better signal processing algorithms for (i) carrier recovery, (ii) channel estimation, and (iii) data detection. The motivation of our work is to ensure accurate signal recovery in the presence of strong ITI by addressing each of these three aspects.
\subsection{Literature Survey}

\subsubsection{Channel Estimation Schemes}

OFDM channel estimation typically consists of obtaining the initial estimates at the pilot locations and then interpolating these estimates over the data subcarriers. The channel interpolation can be done using various methods such as linear interpolation, transform domain techniques, Wiener filtering, etc. \cite{Ozdemir2007Channel}. The initial estimate quality will affect the quality of the interpolated channel. 

The initial channel estimates can be obtained using frequency or time-domain least-squares, the linear minimum mean squared error (LMMSE) approach \cite{van1995channel}, the maximum likelihood (ML) approach \cite{Larsson2001Joint} or the Bayesian MMSE approach \cite{Morelli2001Acomparison}. The ML and time-domain least squares solutions (also called modified least squares or mLS) are found to be equivalent, and the MMSE solution reduces to ML at high signal-to-noise ratios (SNRs) \cite{Ozdemir2007Channel, Morelli2001Acomparison}. Thus, the mLS/ML is a minimum variance unbiased estimator that yields the best performance without knowledge about the channel correlation matrix. The mLS typically uses implicit DFT interpolation but can also work with other interpolation methods.  

Signal detection in the presence of interference requires estimates of the channels of the interfering signals and that of the desired signal. These channels can be estimated by allotting them non-intersecting sets of frequency orthogonal pilot subcarriers like those described in \cite{Karunakaran2011}. However, such designs are not flexible as they would have to reserve pilot sets for the maximum number of BS towers that the user equipment (UE) is likely to detect. The number of such towers is calculated based on inter-site distances and path loss calculations alone. The combination of shadowing, fading, and dynamic interference control techniques can reduce the number of towers that interfere with the UE. A design based on the maximum number of BS towers in a region would be spectrally inefficient as it would require more pilot subcarriers in total. 

The pilot-on-pilot allocation scheme multiplexes the pilot positions of the desired signal and interferers \cite{Karunakaran2011}. The pilots originating from different towers would then be separated with the help of codes rather than frequency orthogonality. Interfering pilots are either suppressed or cancelled out altogether in such an allocation \cite{Abdelkader2020,  Pratschner2015, Pratschner2017}. The 5G-NR standards \cite{3gpp38211}, for example, define the demodulation reference signal (DM-RS) sequence with such a pilot-on-pilot allocation by using orthogonal cover codes (OCCs). The channel is assumed to be static over the time and frequency resources over which the OCC is applied. This assumption makes DM-RS channel estimation suffer from performance degradation in channels with moderate to high selectivity \cite{kong2020mmse,Preethi2022Exploiting}. 

The sounding reference signal (SRS) in 5G also uses pilot-on-pilot arrangement \cite{3gpp38211}. However, the ports use distinct phase-shifted pilot sequences \cite{GeoffreyLi2002} rather than relying on OCC to separate the ports. The phase-shifted pilots result in time-shifted CIR for each port, which can then be separated by windowing the CIR \cite{Tran2018SRS}. However, this approach suffers from an error floor caused due to interpolation error that arises from zero padding in the CIR \cite{Auer2003Channel}.

\subsubsection{Management of CCI and ICI}
Co-channel interference must either be avoided or suppressed  to decode the received signal. The transmitter powers at the different towers can be optimized by solving a system of equations as in \cite{Pietrzyk2003Subcarrier} to minimize the CCI. The idea is to reduce the interference power while keeping the quality at the desired receiver at an acceptable level. While this method utilizes knowledge of channel states of the neighboring cells, recent works like \cite{Xiao2020Reinforcement} have proposed strategies that do not require this knowledge. CCI is suppressed in \cite{necker2006towards} by a combination of interference coordination and beamforming. Beamforming-based techniques use multi-antenna systems and cannot be practically deployed in the sub-1 GHz band. Power control faces limitations in the presence of multiple high-power interferers, as the transmitted power of a tower cannot be reduced without the associated cost of reduced throughput in its home cell. 

Thus, either successive interference cancellation (SIC) or joint detection has to be employed to deal with interference at the receiver \cite{lee2011interference}. When the signal-to-interference ratio (SIR) is very high, the signal can be decoded satisfactorily by ignoring the interference. On the other hand, SIC can be employed when interference power exceeds the signal power by a large margin \cite{Vanka2012Superposition}. A joint detector such as the one described in \cite{lee2011interference, Knopp2012} is the best option when the signal and interference powers are comparable. Such detectors typically utilize a joint loglikelihood ratio (JLLR) metric \cite{vishnu2016joint} that accounts for the interferer constellations and the signal constellation in the minimum distance calculation.

The local oscillators of the different towers seen by the receiver induce frequency offsets with varying signs and magnitudes on the receiver. The total frequency synchronization error is comprised of the error in local oscillator frequencies and the relative Doppler of the mobile UE relative to the BS \cite{Lin2018Synchronization}. The term carrier frequency offset (CFO) will be used in our work to denote this combined error. CFO error is the major contributor\footnote{Inter-Block Interference (IBI) can also contribute to ICI in OFDMA cellular systems. In our work, we assume that there is sufficient cyclic prefix available on the downlink ITI signals from GPS synchronized BS to ensure that the UE determines an IBI-free window.} to inter-carrier interference (ICI) in an OFDM or OFDMA system. 

ICI cancellation literature focuses on removing or suppressing ICI in the time or frequency domains. Frequency domain approaches are limited and more complex as they mostly require reconstructing the ICI to cancel them \cite{kotzsch2009joint}. On the other hand, a time-domain algorithm reduces the CFO before it is formed by the FFT operation \cite{chiueh2008ofdm}. Some methods, like \cite{GangWang2013}, combine time domain and frequency domain approaches. Other methods for multi-transmitter CFO compensation involve singular value decomposition \cite{Hari2017}, lattice filtering \cite{Schlamann2014}, iterative design \cite{Jeon2014Iterative}, etc. Complete compensation for the CFOs is not possible when multiple transmitters are present, as in the case of a reuse-1 system \cite{Jeon2014Iterative}.  

The downlink multi-transmitter CFO problem is distinct from CFO correction in the absence of co-channel interferers for uplink OFDMA links \cite{Morelli2004,Defeng2005}. CCI comes into the picture on the uplink in MIMO systems that use spatial multiplexing \cite{Schellmann2006, kotzsch2009joint}. Based on the downlink CFO estimate, the uplink compensation schemes generally assume some amount of precompensation by the UE \cite{kotzsch2009joint}. UE-specific precompensation is not possible in the downlink as the towers simultaneously cater to multiple UEs. 

The ICI contribution of multiple CFOs in the downlink is studied extensively in Coordinated Multi-Point (CoMP) literature \cite{marsch2011coordinated, Manolakis2011Impairment, Deng2009Correction, yang2015low,
Cavers2010Coordinated}. A receiver in a CoMP system sees multiple towers from nearby cells that transmit desired information as a form of diversity \cite{marsch2011coordinated}. The degradation due to multi-transmitter CFO is studied in \cite{Manolakis2011Impairment}. The ICI that arises from downlink transmitters is an issue that reuse-1 systems share with CoMP schemes. The ICI compensation proposed in \cite{Deng2009Correction} utilizes a time-domain derotation factor that is a nonlinear function of the CFOs. Some corrections are proposed to this derotation factor in \cite{yang2015low}. These methods assume that the difference between the minimum and maximum CFOs is less than half of the subcarrier bandwidth. The CFO correction is shifted to the transmitter side in \cite{Cavers2010Coordinated} by first de-biasing the various CFOs at the receiver and then feeding back the residual to the respective towers. However, \cite{Cavers2010Coordinated} assumes a quasi-static, non-selective channel and a single-carrier system.

\begin{figure*}[!t]
\centering
\includegraphics[scale=0.75]{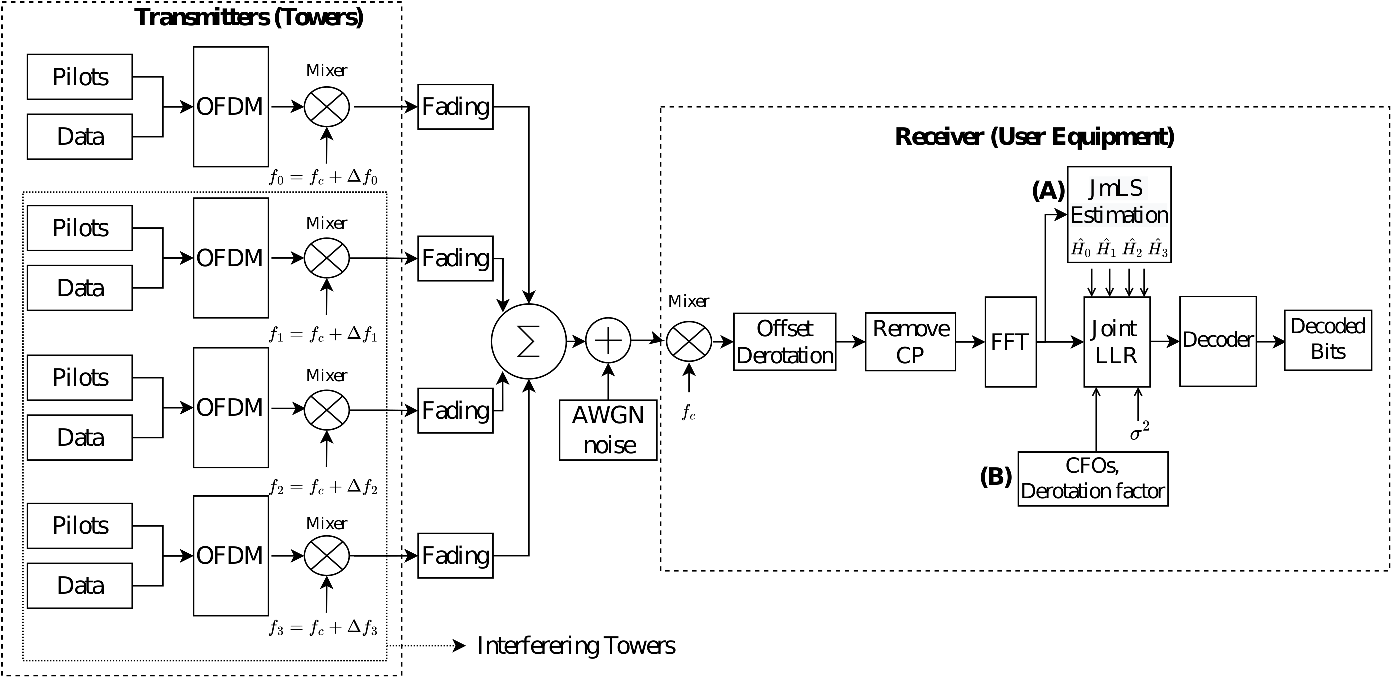}
\caption{Block diagram describing a system with three ITI signals ($M$=4) and the proposed receiver structure.}
\label{Fig1}
\end{figure*}
\subsection{Key Contributions}
This work describes an algorithm that jointly estimates the desired and interference channels in a spectrally efficient manner. The proposed estimator employs the comb-type pilot-on-pilot design strategy described in \cite{Karunakaran2011}. The pilot pattern used at the different towers may be random PSK/QAM sequences. The signal model in this work enables joint estimation by accounting for both interferer and desired signal pilots. The proposed channel estimator is an extension of the mLS estimator \cite{van1995channel}. It will be shown that the combination of mLS and pilot-on-pilot configuration makes this estimator spectrally efficient.  

The time-domain methods, such as those described in \cite{chiueh2008ofdm,Deng2009Correction}, and \cite{yang2015low}, do not entirely compensate for multi-transmitter CFO even when used in conjunction with frequency domain approaches (as seen in \cite{GangWang2013}). The residual CFOs cause phase rotations in the signal and interferer constellations. The interference-aware joint detector proposed in this work tracks and compensates for this phase resulting from the residual CFO.

The key contributions of this work are captured below with reference to the block diagram in Fig. \ref{Fig1}.
\begin{enumerate}
\item We first propose a joint channel estimation algorithm (labeled (A) in Fig. \ref{Fig1}) that can estimate the desired and interference channels in a spectrally efficient manner. This estimator is shown to outperform null-on-pilot and existing pilot-on-pilot-based estimation algorithms \cite{Preethi2022Exploiting,GeoffreyLi2002,Auer2003Channel}. A comparison of the Cramer-Rao lower bound (CRLB) indicates that there is no degradation in using the pilot-on-pilot estimator compared to a conventional null-on-pilot estimator like \cite{van1995channel}. This estimator is shown to be well-defined even when the conventional orthogonal estimator may be under-determined. The proposed estimator is also shown to be a minimum variance unbiased estimator (MVUE).
\item The second contribution is the inclusion of a phase correction factor (labeled (B) in Fig. \ref{Fig1}) in the JLLR-based detection framework discussed in \cite{lee2011interference} and \cite{vishnu2016joint} to work in the presence of distinct CFOs. The phase due to the residual CFO will have to be tracked with dedicated pilots in every OFDM symbol in the absence of this factor. Such pilots are unnecessary overhead and reduces spectral efficiency. Since the LLR is calculated considering the ITI structure, this joint detector manages both interference and CFO errors elegantly using only a single receive antenna.
\end{enumerate}
The proposed approaches for channel estimation and signal detection complement each other because the channel estimates carry information on the distortion caused due to the CFOs to the JLLR-based detector. The joint detector has a simple mechanism to track and compensate for the incremental ``phase ramp" caused due to the residual CFO error. It is shown that the absence of the modification to the JLLR block can lead to an unacceptably high block error rate in the scenarios considered.
\subsection{Related References and Novelty}
The proposed estimator can be considered an extension of the mLS estimator \cite{van1995channel} for the case when the received signal contains CCI. The work in \cite{nguyen2007channel} proposes a similar estimator but requires a specialized training sequence with an entire OFDM symbol reserved for pilots. The proposed estimator requires fewer pilots to estimate the channel and is shown to be MVUE. The present work also identifies the condition for joint estimator superiority over an orthogonal-pilot-based estimator. 

Unlike existing pilot-on-pilot-based methods, the proposed method does not enforce orthogonality using OCC \cite{Preethi2022Exploiting} or phase-shifted pilots \cite{GeoffreyLi2002,Auer2003Channel}. 
Simulation results show that the proposed design is not vulnerable to frequency selectivity. The proposed joint channel estimation scheme allows overlapping multipath delays and does not require interference-free initial channel estimates as in \cite{raghavendra2009interference}. The algorithm proposed in \cite{jeremic2004ofdm} requires the knowledge of the maximum delay spread, whereas our work makes no such assumption. An uplink scenario involving coordinated joint detection of $K$ users by $M$ tower equipment is described in \cite{kotzsch2009joint}. In the downlink model considered here, the UE only needs to decode the message from one of the towers while modeling the interference structure caused by the other ITI terms.

To the best of our knowledge, this is the first work to propose a joint channel estimation framework requiring fewer pilots than its orthogonal counterpart. Unlike an orthogonal pilot system, the proposed estimator needs to be designed for the number of interferers seen by the UE rather than the worst case. The proposed modification to the JLLR expression compensates for the common phase error from the residual CFO and is required when frequency synchronization errors are present. 

This paper is organized as follows. Section II describes the system model under consideration. The key components of the proposed interference-aware receiver are presented in Section III. Section IV presents some numerical simulation results and the results of a link-level simulation. The work is concluded in Section V.
\subsubsection*{Notations}
Bold symbols denote vectors or matrices. Uppercase letters usually indicate the frequency domain, and lowercase letters are for the time domain. A hat on top of a parameter (e.g., $\hat{x}$) will represent an estimate of the parameter $x$. The subscripts $k$ and $m$ are used to denote the $k^{th}$ subcarrier and $m^{th}$ tower, respectively. Finally, the OFDM symbol index shall be denoted by $i$, and $j$ shall be reserved for the unit imaginary number. 

\section{System Model}
An $N$ subcarrier reuse-1 downlink OFDMA system with a cyclic prefix (CP) of $N_{cp}$ samples is considered. Assume that the system operates in the sub-1 GHz UHF band. The link between the base station/tower and the UE is assumed to be single-input single-output (SISO). Although the UE is near $N_T$ base stations/towers, not all of them are visible. 

Consider the situation where only $M<N_T$ towers are typically visible to the UE due to long-term fading, shadowing, interference management or avoidance schemes, etc. A preamble symbol\footnote{The estimate of $M$ is obtained from a banded preamble design, where each of the $N_T$ towers occupy non-overlapping (orthogonal) subcarriers. As explained in Section IV-\ref{SecOCJLLR}, a filter bank can be used at the receiver to detect the presence of interferers.} with $N_T$ bands can be used to estimate $M$. The towers are numbered from $0,1,...M-1$, and the $0^{th}$ tower denotes the one to which the UE is communicating. The other $M-1$ towers transmit co-channel interfering signals since they occupy the same time-frequency resources. With the UE carrier frequency acting as the reference, the CFOs of the $M$ towers are denoted as $\{\Delta f_0, \Delta f_1, ... \Delta f_{M-1}\}$ and the subcarrier bandwidth normalized CFOs are $\{\epsilon_0, \epsilon_1, ... \epsilon_{M-1}\}$ respectively. The data is assumed to be coded. The system is described by the block diagram in Fig. \ref{Fig1} for $M=4$.

A mathematical model for the system can be developed based on the interference-free system model described in \cite{cho2010mimo}. When all UEs have different CFOs, it is observed that the individual constellations of the different transmitters are rotated by an amount corresponding to the respective offset. For the signal belonging to the $m^{th}$ tower, the CFO-induced phase at the $n^{th}$ sample of the $i^{th}$ received OFDM symbol shall be $c_{i,m}(n) = e^{j2\pi\frac{\left(i(N+N_{cp})+n\right)\epsilon_m}{N}}$. Then, the time domain expression for the received OFDM symbol before CP removal is given by 
\begin{equation}
y_i(n)= \sum\limits_{m=0}^{M-1} c_{i,m}(n)\ s_{i,m}(n)+ w_i(n)
\label{SII_E1}
\end{equation}
where $s_{i,m}$ denotes the faded OFDM symbol from the $m^{th}$ tower. Let $x_{i,m}$ be the transmit symbol and $h_{i,m}$ be the channel impulse response (CIR) for the $m^{th}$ downlink signal. Then $s_{i,m}$ is the convolution of $x_{i,m}$ and $h_{i,m}$. The last term $w_i(n) \in \mathcal{N}(0,\sigma^2)$ in \eqref{SII_E1} is an additive white Gaussian noise (AWGN) sample. With the symbol index $i$ omitted for notational convenience, the frequency domain measurements for the $k^{th}$ subcarrier can be written as follows
\begin{multline}
Y[k] = \sum\limits_{m=0}^{M-1} C_m A_m H_{k,m}X_m[k]+ ICI + W[k] \\
C_m := e^{j\left( 2\pi\frac{i(N+N_{cp})+N_{cp} + (N-1)/2}{N}\right)\epsilon_m}, \\ A_m := \frac{sin(\pi \epsilon_m)}{Nsin\left(\frac{\pi \epsilon_m}{N}\right)}\hfill 
\label{SII_E2}
\end{multline}
where $C_m$ is the CFO-induced phase distortion, $A_m$ is the corresponding amplitude distortion, $H_{k,m}$ is the channel frequency response (CFR) and $X_m[k]$ is the modulated symbol placed in the $k^{th}$ subcarrier by the $m^{th}$ tower. The term $ICI$ in \eqref{SII_E2} is the total inter-carrier interference caused by the CFOs. The measurement noise is zero mean and Gaussian with $W[k] \sim \mathcal{N}(0,\sigma^2)$. The phase terms independent of the OFDM symbol index $i$ in the above expression can be absorbed into the frequency response, which is now denoted by $H_{k,m}'$, to yield
\begin{multline}
Y[k] = C_0' H_{k,0}'X_0[k] + \sum\limits_{m=1}^{M-1} C_m'  H_{k,m}'X_m[k] \\ +\sum\limits_{m=0}^{M-1}C_m' I_{l,k}(m) + W[k]
\label{SII_E3}
\end{multline}
where $C_m' := e^{j2\pi\frac{\left(i(N+N_{cp})\right)}{N}\epsilon_m}$ is the symbol index dependent component of the phase distortion. The signal term ($m=0$) and the co-channel interference terms ($m>0$) are shown separately in this equation. Here, $I_{l,k}(m)$ is the ICI term between subcarriers $l$ and $k$ from the $m^{th}$ tower. This term shall be examined in more detail in Section III-B, where the time-domain compensation scheme for ICI is derived. 

Observe that $H_{k,m}'$ in the above expression is an attenuated and phase-rotated version of $H_{k,m}$ in \eqref{SII_E2}. $H_{k,m}'$ can be estimated for every $P^{th}$ OFDM symbol using pilot-aided channel estimation. Starting from the pilot-containing symbol, $i$ takes index values $0, 1, 2, ... P-1$, and again wraps back to $0$ when the channel is estimated again. This is because $C_m'$ also gets absorbed into $\hat{H}_{k,m}'$ during estimation. $C_m'$ produces a ``phase ramp" called CFO-induced common phase error for the OFDM symbols following the pilot-laden symbol. This $\epsilon_m$-dependent phase ramping term $C_m'$ will cause a time-progressive phase shift for the data symbols received after the pilot-laden symbol. This term must be compensated for while modeling the interference-aware detector.
\section{Proposed Interference-Aware Receiver}
The data bits are coded, modulated, and loaded into OFDM subcarriers before transmission. The signals from the different towers are assumed to arrive within the CP duration of the OFDM symbol. This is reasonable at the cell edge since the strong interferers are almost equally distant from the UE. In other words, assuming IBI-free timing recovery at the UE is possible even with strong ITI. The objective is to recover the transmitted data bits from the frequency-domain received symbol \eqref{SII_E3}. The following subsections look at algorithms proposed for joint channel estimation and frequency offset compensation.

\subsection{Joint Channel Estimation}
The choice of the mLS algorithm used to construct the proposed joint channel estimator shall now be justified. In the time domain, the multipath components will be limited to a sparse subset of the first $N_{cp}$ samples of the estimated channel impulse response. Even if the multipath components' exact location is unknown, only $N_{cp}$ samples need to be estimated in the worst case. The samples after $N_{cp}$ can be made zero as the CIR is usually limited to the cyclic prefix length. 

The traditional least squares formulation performs poorly in the presence of noise, as the number of estimates and the number of pilots ($N_p$) are equal. Suppose $N_p$ pilots are used in mLS. Since $N_{cp}$ is usually just a fraction of $N_p$, high-quality channel estimates are obtained as the estimation problem becomes overdetermined. If reduction of pilot overhead is the primary consideration, the number of pilots needed for estimating the channel can be reduced from $N_p$ to $N_{cp}$. The mLS is suitable to estimate interference channels compared to traditional least squares due to this reduction in the pilot requirement. For an $M$ tower system, the number of pilots required reduces from $M N_p$ to $M N_{cp}$. The fourth subsection shows that the proposed joint estimator can further reduce the number of pilots required in interference channels. 

To derive the joint estimator, \eqref{SIII_E6} is rewritten in vector form (note that for the symbols with pilots, $i=0$ and $\tilde{C}_m'=1$) as follows:
\begin{multline}
\mathbf{Y}^{(D)} = \mathbf{X_0} \mathbf{F_{N_{cp}}} \mathbf{\tilde{h}_0}' + \sum\limits_{m=1}^{M-1} \mathbf{X_m} \mathbf{F_{N_{cp}}} \mathbf{\tilde{h}_m}' +\sum\limits_{m=0}^{M-1}\mathbf{\tilde{I}_m} + \mathbf{\tilde{W}} 
\label{SIIIB_E1}
\end{multline}
Here $\mathbf{Y}^{(D)}$ is an $N \times 1$ OFDM received signal vector in the frequency domain, with each entry of the vector corresponding to a particular subcarrier. The superscript $^{(D)}$ and variables with tilde $(\ \tilde{}\ )$ indicate that derotation was done in the time domain to reduce the ICI. More details on derotation schemes shall be provided in the next section. $\mathbf{F}_{N_{cp}}$ is the subsampled DFT matrix consisting of columns from 1 to $N_{cp}$ of the full $N$-point DFT matrix $\mathbf{F}$. $\mathbf{X_m}$ is a diagonal matrix with modulated data or pilots of the $m^{th}$ BS tower as its diagonal elements, and $\mathbf{\tilde{h}_m}'$ is the $N_{cp} \times 1$ CIR vector corresponding to the $N \times 1$ CFR vector $\mathbf{\tilde{H}_{m}'}$. 

The proposed joint estimator allows a comb-type pilot arrangement in the OFDM symbols with pilots. Equation \eqref{SIIIB_E1} can then be rewritten in matrix form as follows. The corresponding matrix dimensions are also provided.
\begin{align}
\begin{split}
&\mathbf{Y}^{(D)}_p = \mathbf{X_p} \ \mathbf{F}_{pN_{cp} M} \ \mathbf{\tilde{h}'}_{N_{cp} M} +\mathbf{\tilde{I}+\mathbf{\tilde{W}}}\quad \quad \quad   \hfill (N_p \times 1)\\
&\mathbf{X_p} = \left[\mathbf{X_{0}} \  \mathbf{X_{1}} \  ... \  \mathbf{X_{M-1}} \right]\quad \quad \quad  \quad \quad  \quad \ \hfill (N_p \times M N_p)\\
&\mathbf{F}_{pN_{cp} M} = \begin{bmatrix}
\mathbf{F}_{pN_{cp}} \quad \mathbf{0} \quad ... \quad \mathbf{0} \\ 
\mathbf{0} \quad \mathbf{F}_{pN_{cp}} \quad ... \quad \mathbf{0} \\
... \quad ... \quad ... \quad ... \\
\mathbf{0} \quad \mathbf{0} \quad ... \quad \mathbf{F}_{pN_{cp}} \\
\end{bmatrix}\quad \quad \  \hfill (M N_p \times M N_{cp})\\
&\mathbf{\tilde{h}'}_{N_{cp} M}=\begin{bmatrix}
\mathbf{\tilde{h}'}_0 \ 
\mathbf{\tilde{h}'}_1 \  
... \ 
\mathbf{\tilde{h}'}_{M-1} \\ 
\end{bmatrix}^T \quad \quad \quad \quad \quad \ \ \hfill (MN_{cp} \times 1)
\label{EJmLSnoOC}
\end{split}
\end{align}

The subscript $_p$ denotes that only pilot subcarriers are selected for those matrices or vectors. The rows corresponding to pilot subcarriers are selected from $\mathbf{F}_{N_{cp}}$ to get $\mathbf{F}_{pN_{cp}}$. The objective now is to solve the following least-squares problem.
\begin{equation}
\min\limits_{\mathbf{\tilde{h}'}_{N_{cp} M}} \|\mathbf{Y}^{(D)}_p - \mathbf{X_p} \ \mathbf{F}_{pN_{cp} M} \ \mathbf{\tilde{h}'}_{N_{cp} M}\|^2
\end{equation}
This minimization can be solved by differentiating the function with respect to $\mathbf{\tilde{h}'}_{N_{cp} M}$ and setting the result equal to zero. 
\begin{equation}
-2 (\mathbf{Y}^{(D)}_p)^H \boldsymbol{X_p} \mathbf{F}_{pN_{cp} M} + 2 {(\mathbf{\tilde{h}'}_{N_{cp} M})}^H \mathbf{F}_{pN_{cp} M}^H \boldsymbol{X_p}^H \boldsymbol{X_p} \mathbf{F}_{pN_{cp} M} = 0.
\label{EqFirstDerivative}
\end{equation}
This can be rewritten as
\begin{equation}
{\mathbf{F}_{pN_{cp} M}}^H \boldsymbol{X_p}^H \boldsymbol{X_p} \mathbf{F}_{pN_{cp} M}\ {\mathbf{\tilde{h}'}_{N_{cp} M}} =  \mathbf{F}_{pN_{cp} M}^H \boldsymbol{X_p}^H \mathbf{Y}^{(D)}_p.
\label{2ndEqFirstDerivative}
\end{equation}
The estimate of $\mathbf{\tilde{h}'}_{N_{cp} M}$ can be obtained by multiplying throughout by the regularized inverse of ${\mathbf{F}_{pN_{cp} M}}^H \boldsymbol{X_p}^H \boldsymbol{X_p} {\mathbf{F}_{pN_{cp} M}}$.
\begin{subequations}
\begin{align}
\mathbf{\hat{h}}_{JmLS} = \mathbf{J}_M \mathbf{Y}^{(D)}_p
\label{JmLSCIR}
\end{align} 
where
\begin{align}
\mathbf{J}_M  =  \left( {\mathbf{F}_{pN_{cp} M}}^H \mathbf{X_p}^H \mathbf{X_p} \mathbf{F}_{pN_{cp} M} + \alpha \mathbf{I}\right)^{-1}{\mathbf{F}_{pN_{cp} M}}^H \mathbf{X_p}^H 
\end{align} 
\end{subequations}
Here, $\alpha > 0$ is a Tikhonov regularization factor \cite{galatsanos1991cross}. When guard subcarriers are present in the OFDM symbol, the reduced DFT matrix loses orthogonality, and the inverse becomes ill-conditioned. The regularization parameter $\alpha$ improves the condition number of this inverse. $\alpha$ can be chosen according to the Hoerl-Kennard-Baldwin formula \cite{hoerl1975ridge,qian2017determination}. As per this formula, the regularization factor would be,
\begin{equation}
\alpha = \frac{MN_{cp}}{{\mathbf{\hat{h}}_{JmLS}}
^H\mathbf{\hat{h}}_{JmLS}}\hat{\sigma}^2
\end{equation}
Here, $MN_{cp}$ is the number of parameters to be estimated. $\hat{\sigma}^2$ is the estimated noise variance. As the regularization factor requires an estimate of the channel impulse response, ${\mathbf{\hat{h}}_{JmLS}}^H\mathbf{\hat{h}}_{JmLS}$ can be initialized to 1 the first time JmLS is applied. The previous CIR estimate can be used in the subsequent JmLS estimations. The initialization value of 1 corresponds to the assumption of a single path with no fading or attenuation. The noise variance estimate is also required for the joint detector, and can be estimated by leaving a few subcarriers blank. 

$\mathbf{J}_M$ shall be called the \textit{joint modified least squares} (JmLS) estimate for the remainder of this paper. The CFRs can be obtained from the JmLS time domain estimate by multiplying each $N_{cp} \times 1$ subvectors in $\mathbf{\hat{h}}_{JmLS}$ with $\mathbf{F}_{N_{cp}}$. This is equivalent to taking an $N$ point FFT of each such subvector in $\mathbf{\hat{h}}_{JmLS}$. 
\subsubsection{CRLB of the joint CIR estimate}
The pilots are assumed to be equally spaced in the frequency domain from band-edge to band-edge to derive the CRLB. The residual ICI term $\mathbf{\tilde{I}}$ is merged with the noise term $\mathbf{\tilde{W}}$. Under these assumptions, following the steps given in Appendix \ref{Appendix:CRLBJmLS}, CRLB of the total variance of an unbiased joint estimator of $\mathbf{\tilde{h}'}_{N_{cp} M}$ is given by
\begin{equation}
CRLB_{tot}(\mathbf{\hat{h}_J}) = Tr\left( ({\mathbf{F}_{pN_{cp}M}}^H \mathbf{F}_{pN_{cp}M})^{-1}  \right)\sigma^2
\label{ECRLBJmLSQAMpilMain}
\end{equation}
where $Tr(.)$ denotes the trace of a matrix. If the pilot spacing is assumed to be a power of 2, as given in Appendix \ref{Appendix:CRLBJmLS}, then 
\begin{equation}
CRLB_{tot}(\mathbf{\hat{h}_J}) = \frac{M N_{cp}}{N_p^{(J)}} \sigma^2
\label{CRLBJmLSMain}
\end{equation}
where the number of joint estimator pilots $N_p^{(J)} = N/2^i$, $i$ being a non-negative integer. Equation \eqref{CRLBJmLSMain} is only approximate since the inverse will not be a pure identity matrix in the presence of guard subcarriers. Equation \eqref{ECRLBJmLSQAMpilMain} can be used in such cases. 
\subsubsection{JmLS as a Minimum Variance Unbiased Estimator (MVUE)}
The JmLS estimator can be shown to be unbiased by simply substituting  \eqref{EJmLSnoOC} into \eqref{JmLSCIR} and performing the expectation operation. The MVUE property of an unbiased estimator can be proven if its MSE is shown to be equal to the CRLB. The total MSE of the JmLS estimates is now computed as follows:
\begin{equation}
MSE_{_\Sigma} = Tr\left(E\left[(\mathbf{\hat{h}}_{JmLS} - \mathbf{\tilde{h}'}_{N_{cp} M}) (\mathbf{\hat{h}}_{JmLS} - \mathbf{\tilde{h}'}_{N_{cp} M})^H \right]\right)
\label{EqMSE}
\end{equation}
The following expression is obtained by applying the steps given in Appendix \ref{Appendix:MVUE}. For unit amplitude pilots,
\begin{equation}
MSE_{_\Sigma}= Tr \left(\left( {\mathbf{F}_{pN_{cp}M}}^H  \mathbf{F}_{pN_{cp}M} + \alpha \mathbf{I}\right)^{-1} \right)\sigma^2
\label{MSETotJmLS}
\end{equation}
Equation \eqref{MSETotJmLS} is the same as the CRLB expression \eqref{CRLBJmLSMain} except for the regularization factor. When pilots are band-edge to band-edge, and the pilot spacing is assumed as per the CRLB derivation, regularization is unnecessary. For $\alpha=0$, \eqref{MSETotJmLS} becomes equal to \eqref{ECRLBJmLSQAMpilMain}. Thus the proposed JmLS estimator in \eqref{JmLSCIR} achieves CRLB and is an MVUE as well as an efficient estimator. 

\subsubsection{CRLB of Orthogonal CIR estimation scheme}
The CRLB of the joint estimator can now be compared to the mLS algorithm with a null-on-pilot allocation. An orthogonal pilot pattern with $N_p^{(O)}$ pilots per BS tower can be compared to the JmLS pilot-on-pilot pattern with a total of $N_p^{(J)}$ pilots. This orthogonal scheme can estimate CIR using mLS for each tower. This algorithm shall be called \textit{orthogonal modified least squares} (OmLS) to distinguish it from the joint estimation scheme. The OmLS estimate is described by 
\begin{subequations}
\begin{gather}
\mathbf{\hat{h}}_{OmLS} = \mathbf{O}_m \mathbf{Y}^{(D)}_{p,m}\label{OmLSCIR}
\\
\mathbf{O}_m =  \left( \mathbf{F}_{pN_{cp}}^H \mathbf{X_m}^H\mathbf{X_m} \mathbf{F}_{pN_{cp}} + \alpha \mathbf{I}\right)^{-1}\mathbf{F}_{pN_{cp}}^H \mathbf{X_m}^H \label{OmLSCIR2}
\end{gather} 
\end{subequations}
$\mathbf{Y}^{(D)}_{p,m}$ is the vector consisting of only those subcarriers in $\mathbf{Y}^{(D)}_{p}$ that correspond to the $m^{th}$ tower as per the null-on-pilot scheme. The CRLB of OmLS is the special case of JmLS with $M=1$, except for a pilot boosting factor which shall be explained shortly. The CRLB of the total variance of the estimate for the $m^{th}$ user is then
\begin{equation}
CRLB(\mathbf{\hat{h}}_{orth,m}) = Tr\left( ({\mathbf{F}_{pN_{cp}}}^H E\left[{\mathbf{X_m}}^H {\mathbf{X_m}}\right] {\mathbf{F}_{pN_{cp}}})^{-1}  \right)\sigma^2
\label{ECRLBOmLS}
\end{equation}
The number of BS towers $M$ seen by the UE changes with time and location. OmLS can then only be designed for $N_{T}>M$ towers, the maximum number a UE is likely to see at the cell edge. Since $(N_{T}-1)N_{p}^{(O)}$ subcarriers are nulls in OmLS, the power of the $N_{p}^{(O)}$ pilot subcarriers can be boosted by a factor $\beta = N_{T}$. This means $E\left[{\mathbf{X_m}}^H {\mathbf{X_m}}\right] = \beta\mathbf{I}$. The expression in \eqref{ECRLBOmLS} can be further simplified in terms of the pilot boosting factor and the number of OmLS pilots.
\begin{align}
CRLB(\mathbf{\hat{h}}_{orth,m}) & = Tr\left( (\beta N_{p}^{(O)}  \mathbf{I})^{-1}  \right)\sigma^2 = \dfrac{N_{cp}}{\beta N_{p}^{(O)}} \sigma^2
\label{ECRLBOmLS2}
\end{align}
JmLS and OmLS are allotted the same total number of pilot subcarriers to enable a fair comparison. That is, for $N_{T}$ BS towers, $N_{p}^{(O)} = N_{p}^{(J)}/N_{T}$. The aggregate CRLB for the CIRs of all the users is obtained by substituting the value of $\beta$.
\begin{align}
\sum_m CRLB(\mathbf{\hat{h}}_{orth,m}) = \dfrac{M N_{cp}}{\ N_{p}^{(J)}} \sigma^2
\end{align}
This CRLB is identical to the CRLB derived for JmLS in \eqref{CRLBJmLSMain} and it shows that there is no degradation by going for a joint estimator instead of using orthogonal pilots. The next subsection shows that JmLS has an advantage over OmLS in terms of the number of pilots needed.

\subsubsection{Condition for JmLS superiority over OmLS}
\label{JmLSSupSec}
The model in \eqref{EJmLSnoOC} can be rewritten as a system of equations.
\begin{equation}
\mathbf{X_p}^H \mathbf{Y}^{(D)}_p = \mathbf{X_p}^H \mathbf{X_p} \ \mathbf{F}_{pN_{cp} M} \ \mathbf{\tilde{h}'}_{N_{cp} M} +\mathbf{\tilde{I}+\mathbf{\tilde{W}}}
\label{EXHermYp}
\end{equation}
This is a system with $M N_p^{(J)}$ equations and $M N_{cp}$ unknowns. As $N_p^{(J)} = N_{T} N_p^{(O)}$, JmLS would be solvable when $N_p^{(O)} N_{T}> N_{cp}$. When $N_p^{(O)} < N_{cp}$, the OmLS estimator is underdetermined as it would have fewer equations than the number of unknowns. This means that the matrix inverted in \eqref{OmLSCIR2} will not be full rank. Thus, JmLS will be able to perform better than OmLS whenever the following set of inequalities hold true.
\begin{align}
\begin{split}
N_p^{(O)} N_{T}/N_{cp}&>1\\
N_p^{(O)}/N_{cp}&<1
\end{split}
\label{JmLSSup}
\end{align}
It follows that JmLS can work with fewer pilots than OmLS when multiple towers are present.
\subsubsection{Comparison of Complexity}
The estimator matrix $\mathbf{J}_M$ in \eqref{JmLSCIR} is an $M N_{cp} \times N_p^{(J)}$ matrix. Both $\mathbf{X_p}$ and $\mathbf{F}_{pN_{cp} M}$ are required to calculate this matrix. These are computed with the individual pilot matrices $\{\mathbf{X_m}\}$, and the subsampled DFT matrix $\mathbf{F}_{pN_{cp}}$. Since the transmitted pilot sequence and their locations are known at the receiver, $\mathbf{J}_M$ can be precalculated and stored. $\mathbf{J}_M$ should be computed for different $M$ values and the various tower combinations that the receiver is likely to see. The appropriate estimator matrix can then be chosen by determining which $M$ out of the $N_T$ possible bands in the banded preamble (as explained in footnote 3 and the simulation section) contain signals of significant strength. The OmLS estimator matrices for the $N_T$ towers can also be precalculated and stored using similar logic. 

The OmLS and JmLS estimations reduce to matrix multiplications by employing the precalculated estimator matrices at the receiver. A well-determined OmLS-based estimator requires $N_p^{(O)}=N_{cp}$ subcarriers per tower and $N_T N_{cp}$ subcarriers in total. However, JmLS can work with slightly more than $N_{cp}$ pilots. The total number of pilots in OmLS and JmLS are kept equal in this work for a fair comparison of the schemes. This means that JmLS is well-determined and will perform well, while OmLS is underdetermined. For this design, the complexity of JmLS in terms of complex multiplications is $O(M N_{cp} N_p^{(J)})$. The corresponding complexity of OmLS to estimate the CIR for all the channels is  $M\times O(N_{cp} N_p^{(O)}) = O(M N_{cp} N_p^{(O)})$. 

In the case described above, $N_p^{(J)} = N_T N_p^{(O)} = N_{cp}$ which implies $N_p^{(O)} = N_{cp}/{N_T}$. Although OmLS is less complex than JmLS, the estimates are meaningless as $N_p^{(O)}<N_{cp}$. Thus, JmLS requires more complex multiplications to produce reliable channel estimates. On the other hand, OmLS is less complex but underdetermined. This higher complexity can be justified as the JmLS estimator can work with fewer pilots than OmLS. If the total number of pilots of both estimators are fixed such that the equations are well determined in either case, $N_p^{(J)}$ becomes equal to $N_{cp}$ and $N_p^{(O)}=N_{cp}$. In this case, OmLS and JmLS would have the complexity of the order of $O(M N_{cp}^2)$. However, OmLS would require $N_T$ times more pilots in total.

\subsection{Frequency offset compensation}

\subsubsection{Time-domain ICI Compensation Scheme}
\label{S3a}

Time-domain compensation is a computationally simple first step towards reducing the ICI power at the receiver. The CFOs should be estimated at the receiver as their true values are unknown. This work estimates CFOs using a banded preamble, as explained in the simulation section. These estimates are then used to perform offset correction in the time domain. Some of the choices for the compensation factor are:
\begin{enumerate}
\item Desired signal CFO \cite{chiueh2008ofdm}.
\item A function of the arctan of the CFOs as proposed in CoMP literature \cite{Deng2009Correction, yang2015low}.
\item Mean or weighted mean of the CFOs can also serve as a near-optimal derotator as proposed in this work.
\end{enumerate}

In CCI-free systems, ICI is mostly removed with the help of the time-domain scheme described in \cite{chiueh2008ofdm}. In this algorithm, the received time domain signal \eqref{SII_E1} is multiplied with the conjugate of $c_{i,0}(n)$ to remove the effect of CFO. This is equivalent to ``de-rotating" the signal that was rotated by the CFO, and hence this approach is called signal derotation (SD). Some amount of ICI shall remain as an estimate of $\epsilon_0$ is used in practice. 

Although the optimal time-domain derotation factor for a co-channel interference system has not been explored in the literature, there are several studies made in the related field of downlink CoMP. These studies have been outlined in the literature survey. In \cite{Deng2009Correction}, the derotator proposed is the arctan of the ratio of the weighted sum of sines and cosines of the normalized CFOs. A later study \cite{yang2015low} corrected this expression and found that  even for the case of $M=2$, certain cross terms should be accounted for within the arctan expression. It is shown that for $M=2$ and similar fading conditions, the derotator reduces to the mean of the normalized CFOs. For dissimilar fading conditions, the derotator CFO shall move towards the CFO of the signal with the larger fading factor, i.e., lower path loss. This corrected expression has not been derived for $M>2$. 

In this work, a simple linear function of the CFOs is shown to be a near-optimal derotator by  using a slightly different set of approximations than the ones used in \cite{Deng2009Correction} and \cite{yang2015low}. The proposed derotator is a weighted mean of the normalized CFOs and shall be called the weighted mean derotator (WMD). The formulation of the optimization problem and the steps to arrive at this solution is given in Appendix \ref{Appendix:Derotation}. When the user equipment is near the cell edge, the desired and interferer channels have similar path loss. The weighting factors of the CFOs of the different towers become equal in such scenarios, and the WMD reduces to a simple mean derotator (MD). Their expressions are given by

\begin{align}
\epsilon_{wmd} = \frac{\sum\limits_{m=0}^{M-1}P_m \epsilon_m }{\sum\limits_{m=0}^{M-1} P_m}
\end{align}
and 
\begin{align}
\epsilon_{md} = \frac{1}{M}\sum\limits_{m=0}^{M-1} \epsilon_m 
\end{align}

As mentioned at the start of this subsection, the normalized CFOs used in the MD and WMD expressions are replaced by their estimates obtained using a preamble described in section \ref{SimResults}-\ref{SecOCJLLR}. The simulation section explains how the weights used in the WMD formulation can also be obtained from this preamble. It will be shown through simulations that there is no loss in terms of block-error rate (BLER) by using the mean derotator over WMD. 

ICI compensation can be carried out using either the proposed derotation factor or the arctan-based derotator. For any derotation factor $\epsilon_d$, the time-domain derotation is given by
\begin{equation}
y_{D,i}(n) = e^{-j2\pi\frac{\left(i(N+N_{cp})+n\right)\epsilon_d}{N}} y_i(n)
\label{SIII_MD}
\end{equation}
Here $i$ is the OFDM symbol index. $i$ is 0 for OFDM symbols in which channel estimation takes place (pilot-containing symbols). For symbols that do not contain pilots, $i$ increments by 1 for every symbol until the next pilot-containing symbol is reached. In place of \eqref{SII_E3}, the following new equation is obtained after derotation
\begin{multline}
Y_D[k] = \tilde{C}_0' \tilde{H}_{k,0}'X_0[k] + \sum\limits_{m=1}^{M-1} \tilde{C}_m'  \tilde{H}_{k,m}'X_m[k] \\ +\sum\limits_{m=0}^{M-1}\tilde{C}_m' \tilde{I}_{l,k}(m) + \tilde{W}[k]
\label{SIII_E6}
\end{multline}
$Y_D$ is the frequency domain received signal that has undergone time-domain derotation and $\tilde{C}_m' := e^{j2\pi\frac{i(N+N_{cp})}{N}(\epsilon_m-\epsilon_d)}$. The tilde in the above terms indicate that $\epsilon_m$ has been replaced with $\epsilon_m-\epsilon_d$. 

\subsubsection{Frequency offset compensation for joint detector}

A maximum likelihood detector that compensates for the effect of residual CFO phase ramp can be derived using the model described in \eqref{SIII_E6}. The joint detector can be formulated while including the effect of residual CFO as
\begin{equation}
LLR_{0,\lambda,k} = ln \left( \dfrac{P\left(b_\lambda \left( X_0[k]\right) = 1\; \middle| \;Y_D[k],\mathbf{\tilde{C}}', \mathbf{\tilde{H}}_k'\right)}{P\left(b_\lambda \left( X_0[k]\right) = 0\; \middle| \;Y_D[k],\mathbf{\tilde{C}}', \mathbf{\tilde{H}}_k' \right)} \right).
\label{LLRproblem}
\end{equation}
\noindent where the notation $b_\lambda (.)$ denotes the $\lambda^{th}$ bit of the modulated symbol on the $k^{th}$ subcarrier. The probabilities in the numerator and denominator would be Gaussian, assuming that the residual ICI does not much distort the noise distribution. Then, the Bayes theorem can be applied while assuming equal prior probabilities for the constellation points. Following the steps given in Appendix \ref{Appendix:JLLR}, the Max-Log-MAP approximation \cite{robertson1997optimal} for the solution is given by
\begin{multline}
LLR_{0,\lambda,k} \approx  \min\limits_{\substack{X_0 \in \mathbf{X}_0^{\lambda_1}, \\X_m \in \mathbf{X_m}\\m \neq 0}}  \frac{1}{\sigma^2}\| Y_D[k] - \tilde{C}'_0 \ \tilde{H}_{k,0}X_0[k] \\ - \sum\limits_{m=1}^{M-1} \tilde{C}'_m \  \tilde{H}_{k,m}X_m[k] \|^2 \\ 
 - \min\limits_{\substack{X_0' \in \mathbf{X}_0^{\lambda_0},\\X_m[k] \in \mathbf{X_m}\\m \neq 0} } \frac{1}{\sigma^2}\|  Y_D[k] - \tilde{C}'_0 \ \tilde{H}_{k,0}X_0'[k]  \\- \sum\limits_{m=1}^{M-1} \tilde{C}'_m \  \tilde{H}_{k,m} X_m[k] \|^2
 \label{eqOCJLLR}
\end{multline}
The solution is a modified version of the joint log-likelihood ratio (JLLR) \cite{vishnu2016joint} detector and shall be called offset-corrected JLLR (OC-JLLR). The residual ICI is considered a part of the effective noise in the above approximation. The LLR expression  includes the phase ramp term $\tilde{C}_m'$ that tracks the phase change in the data symbols caused due to CFO. It will be shown that block error rate flooring occurs due to CFO-induced ICI in the absence of this term in the JLLR expression. Equation \eqref{eqOCJLLR} can be interpreted as the minimum distance receiver for a ``super-constellation" derived from the desired and interference signals. 


\section{Simulation Results and Discussion}
\label{SimResults}
\subsection{Time-domain ICI Compensation Scheme}
\noindent The extent of residual ICI after compensation depends on the time-domain ICI compensation technique used. Numerical simulations are performed to compare the average residual ICI power of schemes such as signal derotation (SD) \cite{chiueh2008ofdm}, arctan-based derotation (TD) \cite{Deng2009Correction, yang2015low}, the proposed mean and weighted mean derotation (MD, WMD), and a brute-force search for the optimal derotator. The simulation parameters are described in Table \ref{Table1}.

The amount of CFO experienced by signals from various towers can be quantified as follows. The oscillator at the UE has significantly lower accuracy requirements than the BS \cite{Lin2018Synchronization}. Typical numbers for oscillator accuracy in 5G are 10 parts per million (ppm) of the carrier frequency at the UE \cite{Yuichi2020Epson} and as low as 0.1 ppm for base stations that cover a moderate area \cite{Lin2018Synchronization}. It is then evident that the CFO will have the same sign for all the ITI terms. The CFO would be within $10 \pm 0.1$ ppm per the tightest specifications in 5G-NR. 

\begin{figure}
\includegraphics[width=1\columnwidth]{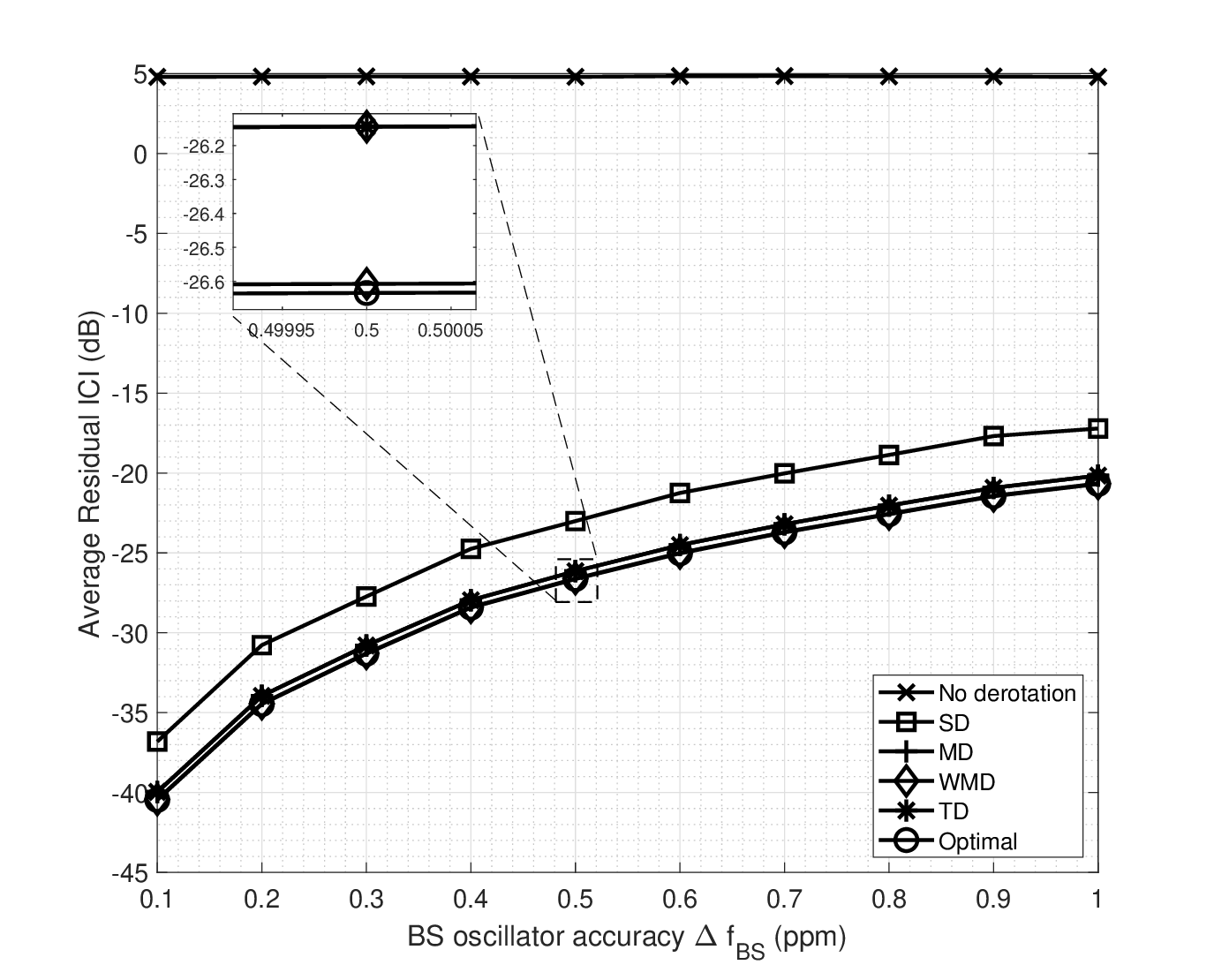}
\caption{Average residual ICI power vs. BS CFO accuracy for various derotation factors}
\label{FigICI1}
\end{figure}

In the first scenario considered, the normalized CFOs are chosen from the interval $[\epsilon_{min},\epsilon_{max}]$. Here, $\epsilon_{min} = \frac{(\Delta f_{UE} - \Delta f_{BS}) f_c}{\Delta f_{sc}}$ and $\epsilon_{max} = \frac{(\Delta f_{UE} + \Delta f_{BS})f_c}{\Delta f_{sc}}$. $\Delta f_{sc}$ is the OFDM subcarrier spacing, $\Delta f_{UE}$ is the local oscillator error of the UE in ppm, $\Delta f_{BS}$ is the local oscillator error of the BS in ppm, and $f_c$ is the carrier frequency. As $\Delta f_{UE}$ is at least an order lower than $\Delta f_{BS}$ (as UE typically has poorer quality equipment), both $\epsilon_{min}$ and $\epsilon_{max}$ are positive. Thus, the normalized CFO varies over a relatively small range. For $\Delta f_{UE}=10$ ppm and $\Delta f_{BS} = 1$ ppm, the normalized CFO ranges between $0.3$ and $0.36$. Thus, the CFOs will have the same sign. Time domain compensation is most effective in this scenario since the common frequency offset can be removed via derotation. The average residual ICI corresponding to CFO ranging from $10 \pm 0.1$ ppm to $10 \pm 1$ ppm are plotted in Fig. \ref{FigICI1}. The case when derotation is absent is also plotted. 

\begin{table} 
        \centering
        \caption{Numerical Simulation Parameters}
        \begin{tabular}{l l}
            \toprule
            Parameter & Value \\
            \midrule 
			 Carrier Frequency & 500 MHz\\
            Max. no. of Towers ($N_T$)&8\\       			 
            No. of Towers ($M$)&4\\       
            Relative received powers& [0 3 -3 0] dB\\    
            CFO range& 10 $\pm$ 0.1 ppm  to\\
            \ & 10 $\pm$ 1 ppm\\
            Subcarrier Bandwidth & 15 kHz\\
            FFT Size & 2048\\
			Channel Model & TDL-B\\
			RMS Delay Spread & 300 ns\\
			CP length ($N_{cp}$) & 72 samples\\
            \bottomrule
        \end{tabular}
        \label{Table1}
\end{table}
It can be seen that the residual ICI is very high ($5$ dB) in the absence of derotation. Among the derotation techniques, signal derotation performs worst. WMD is very close to the optimum derotation found by brute force search. It outperforms the arctan-based derotator proposed in \cite{Deng2009Correction} by approximately 0.5 dB. The mean derotator and the arctan-based derotator have very similar performances.

\begin{figure}
\includegraphics[width=1\columnwidth]{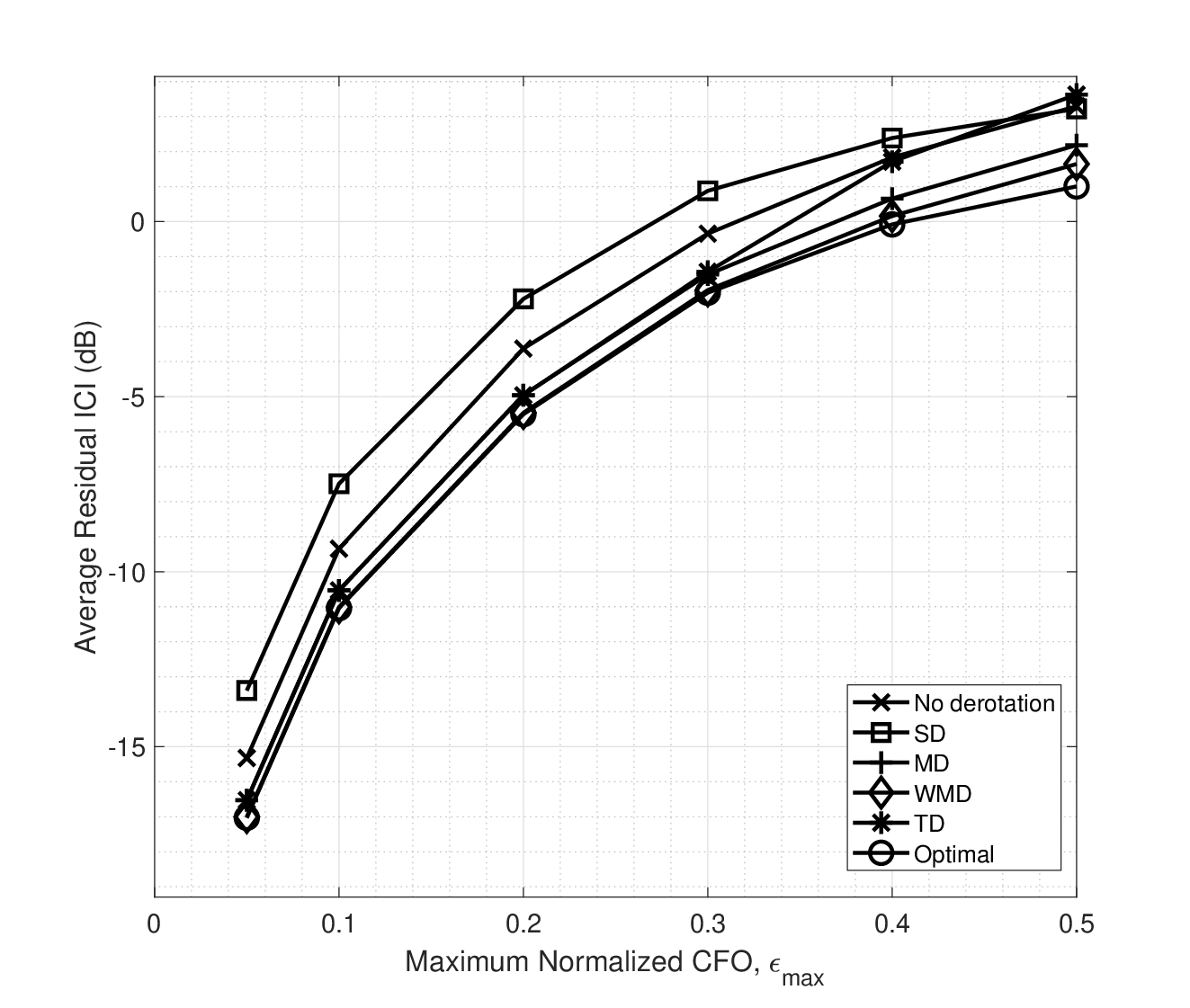}
\caption{Average residual ICI power as a function of maximum normalized CFO}
\label{FigICI2}
\end{figure}

The carrier frequency offsets are randomly drawn from $[-\epsilon_{max},\epsilon_{max}]$ when no assumptions are made regarding the transmitter (base station) and receiver (UE) oscillator accuracy. In this case, the average residual ICI power is plotted in Fig. \ref{FigICI2} for various values of $\epsilon_{max}$. Since the CFOs are randomly distributed around 0, the mean value will also be close to zero. This means that the extent of offset compensation via derotation is limited. The level of residual ICI is seen to be higher in this scenario. It is seen that the case without any ICI compensation (no derotation) performs better than the signal derotation technique. This is because the mean value of the CFOs is near zero and `no derotation' is the same as derotating with zero. Depending on the CFO of the desired signal, SD could perform the derotation with the minimum or maximum value among the CFOs. This kind of derotation can further amplify the ICI rather than reduce it. Hence it is clear that in such scenarios, it is better not to perform derotation than derotating with the desired signal CFO.

In Fig. \ref{FigICI2}, WMD is seen to be near-optimal even for a random distribution of CFO. The arctan-based derotator (TD) \cite{Deng2009Correction} performs well for $\epsilon_{max}<0.3$ but degrades for larger $\epsilon_{max}$. This degradation is because the assumptions made during the derivation of this technique require that $\epsilon_{max}-\epsilon_{min}\le 0.5 \implies \epsilon_{max}\le 0.25$. The WMD uses a different set of assumptions and approximations (as given in Appendix \ref{Appendix:Derotation}) and is seen to perform close to the optimal derotator in practice.
\subsection{JmLS Channel Estimation Performance}
\begin{figure}
\includegraphics[width=1\columnwidth]{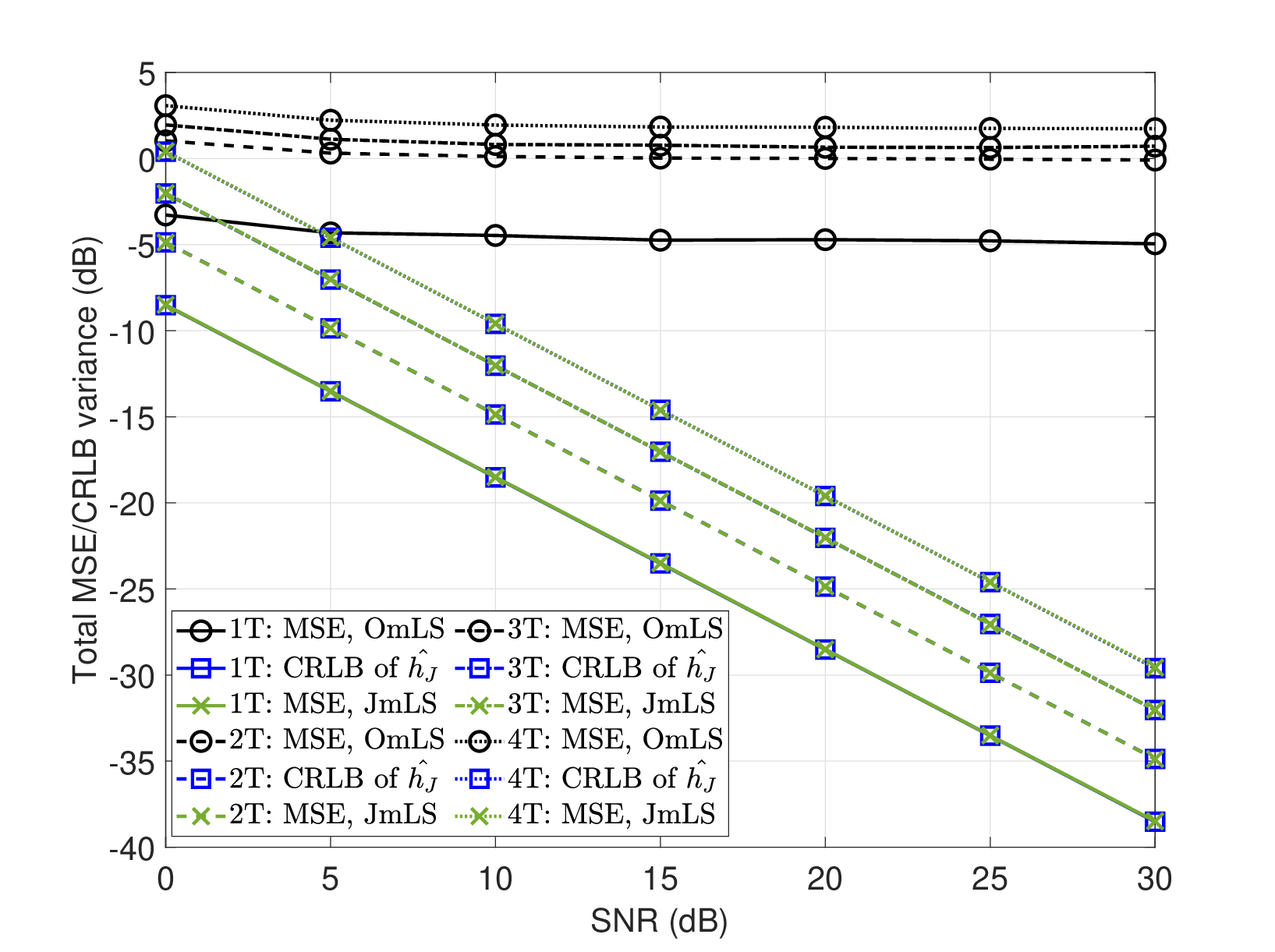}
\caption{Total MSE of JmLS and OmLS CIR estimates plotted along with CRLB for the joint estimator. In the legend, $m$T indicates $M=m$ interfering signals. }
\label{FigRB1}
\end{figure}

The MSE performance of the JmLS channel estimator is studied in this section. First, the total MSE of the CIR for all the towers is calculated using link-level simulations. This is plotted along with the theoretical CRLB on the CIR of the joint estimator given by \eqref{ECRLBJmLStot}. $N_T = 8$ and $M \le 4$ is considered for this simulation. The JmLS channel estimator uses $N_p^{(J)} = 512$ band-edge to band-edge pilots. The corresponding OmLS design uses $N_p^{(O)} = 64$ pilots per tower as the system has to be designed for eight towers. The OmLS estimator is underdetermined since $N_p^{(O)} < N_{cp}$ but the JmLS estimator is well conditioned. Even though the CRLB of the OmLS estimator is not defined as the equations are underdetermined, the MSE can still be found.  The results in Fig. \ref{FigRB1} show that the JmLS estimator MSE coincides with the CRLB for the joint estimator and decreases linearly with increasing SNR. On the other hand, the MSE of the OmLS estimator does not improve with SNR because the number of pilots are insufficient.

The following results are based on a calibrated link-level simulation of an OFDMA framework with parameters shown in Table \ref{Table2}. Some parameter values have changed from Table \ref{Table1}. The UE now occupies only a fraction of the total usable subcarriers. 

Estimators are now compared in terms of the average per-subcarrier MSE of the CFR. JmLS is compared with the OmLS estimator and two existing pilot-on-pilot estimation algorithms. These are the robust MMSE algorithm \cite{Hoeher19972D} which can be used to refine and interpolate channel estimates obtained using 5G DM-RS, and the windowed DFT-based algorithm \cite{Tran2018SRS, GeoffreyLi2002, Auer2003Channel} used to extract channel information from phase-shifted pilots employed in 5G SRS. 

The literature survey explains that the DM-RS uses pilot-on-pilot allocation by spreading pilot subcarriers across time and frequency. OCCs are used to spread the pilots and multiplex DM-RS ports in the same set of pilot subcarrier resources. These ports may correspond to the channels of the spatially multiplexed layers of the same link or the channels of different links. In the reuse-1 scenario, the desired and interferer channels are assigned to distinct DM-RS ports. The first step in the receiver is to perform pilot `despreading' and least-squares estimation as described in \cite{Preethi2022Exploiting}. These estimates are subsequently refined and interpolated using robust MMSE or Weiner filtering as described in \cite{Hoeher19972D}. The DMRS-MMSE estimator is assumed to know the true value of delay spread and SNR for calculating the Weiner filter weights. 

A pilot sequence based on double symbol type-II DM-RS \cite{3gpp38211} is used as one of the benchmarks to compare the performance of the proposed JmLS scheme. One of the three CDM groups is reserved for pilots to accommodate $M=4$ towers. If one DM-RS additional position\footnote{The terms such as `type-II', `CDM group', `additional position', etc. are specific to 5G. More details can be found in \cite{3gpp38211} and \cite{Preethi2022Exploiting}.} is considered, this amounts to four pilot-containing symbols per slot. A 20 PRB allocation  would have $20 \text{ PRB } \times 12 \times \frac{1}{3}\text{ pilot subcarriers per PRB}= 80$ pilot subcarriers per pilot-containing symbol. The pilot overhead is the same as that for JmLS in this configuration of DM-RS. 

The second pilot-on-pilot scheme in the literature is used in 5G SRS channel estimation. Phase-shifted sequences are used to separate channel impulse responses in time so that they can be windowed out in the receiver \cite{Auer2003Channel}. This approach causes interpolation errors due to zero padding in practical non-sample-spaced channels. In addition to OmLS and DMRS-MMSE, phase-shifted pilots coupled with windowed DFT-based channel estimation shall also be compared with the JmLS scheme. This approach shall be labeled as SRS-DFT, although it is not specific to SRS. Once again, the total number of pilots shall be the same as JmLS.

\begin{table} 
        \centering
        \caption{OFDMA Link-Level Simulation Parameters}
        \begin{tabular}{l l}
            \toprule
            Parameter & Value \\
            \midrule
            Time of Flights (for $M=4$) & [0 150 300 450] ns\\
			 Used Subcarriers & 240 (20 PRBs)\\
			 No. of JmLS Pilots  $(N_p^{(J)})$ & 80\\
			 No. of OmLS Pilots  $(N_p^{(O)})$ & 10\\			 
            \bottomrule
        \end{tabular}
        \label{Table2}
\end{table}

\begin{figure}
\includegraphics[width=1\columnwidth]{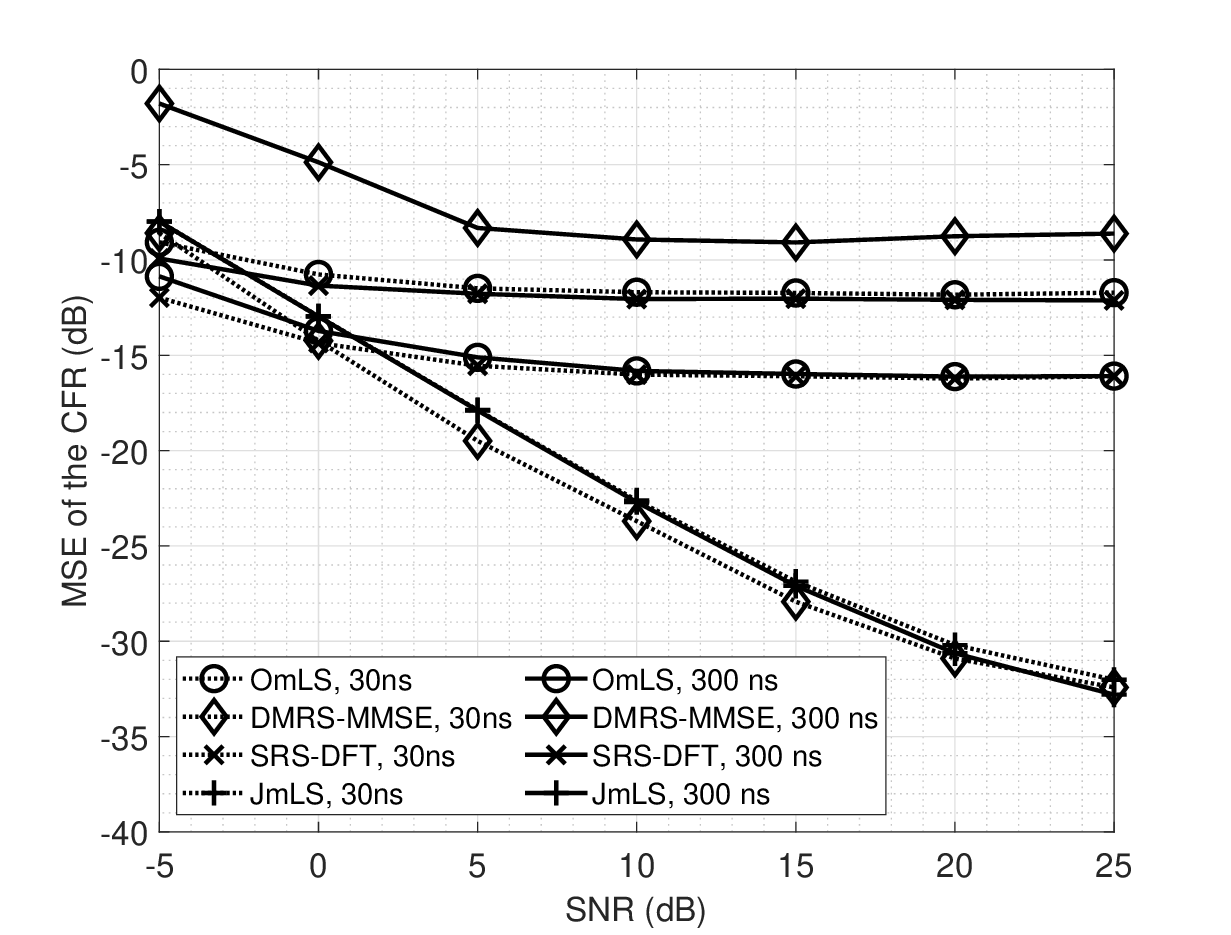}
\caption{MSE performance comparison of JmLS with existing schemes}
\label{FigRB2}
\end{figure}

The result in Fig. \ref{FigRB2} compares the performance of JmLS with that of OmLS, DMRS-MMSE, and SRS-DFT. For a 30 ns delay spread, DMRS-MMSE is marginally superior to the proposed JmLS estimator. However, at 300 ns delay spread, it is clear that DMRS-MMSE floors with a high MSE while the JmLS performance does not change. That is, the DMRS-MMSE is not suitable for moderate to highly frequency selective channels. The DM-RS scheme cannot track the frequency selectivity well because it will effectively yield only $2$ estimates per PRB. JmLS, with $80$ pilots in $240$ subcarriers, gives $4$ estimates per PRB. JmLS has twice the number of estimates and hence does not floor at high delay spreads. 

The OmLS MSE floors at high SNRs because the number of pilots per tower is the total number of pilots divided by $N_T$, which is 8 in this case. Thus, the number of pilots per tower becomes insufficient to estimate the channel with good quality when the allocation is orthogonal. JmLS MSE does not floor because it depends on the visible number of towers rather than the maximum number of towers, i.e., $M$ rather than $N_T$. OmLS floors at a lower MSE than DMRS-MMSE as mLS-based schemes that estimate the channel in time-domain have a natural advantage due to the rejection of the noise samples beyond the cyclic prefix length. 

The SRS-DFT scheme also floors as the TDL-B channel model is non-sample-spaced. The MSE performance of the SRS-DFT scheme shown in Fig. \ref{FigRB2} is similar to that found in \cite{Auer2003Channel}. It is seen that JmLS is the only one among the four schemes whose performance does not floor, irrespective of the delay spread used. It is also spectrally efficient as it provides the best estimation quality with the least number of pilots.

\subsection{Link-Level Simulation Performance}
\label{SecOCJLLR}
The proposed receiver is now evaluated in terms of coded BLER performance. LDPC codes of rate 1/2 are used for this simulation. The signal modulation from each tower is assumed to be known at the receiver. When this information is unknown, modulation order classification techniques such as those described in \cite{Gomaa2016} and \cite{Sai2020} could be used. The noise variance, $\sigma^2$, can be estimated either from a few subcarriers left purposefully blank or using guard subcarriers if the adjacent channel interference is low. The first approach is used in the simulations presented in this paper. 

The link-level simulator follows the frame structure described in 5G standards \cite{3gpp38211} with \mbox{15 kHz} subcarrier spacing. Every frame has ten slots, and every slot has 14 OFDM symbols. The simulations use a time-varying channel with Doppler corresponding to \mbox{10 kmph} for the low-mobility scenario and 100 kmph for the high-mobility scenario. The channel is estimated in the $1^{st}$, $5^{th}$, $9^{th}$ and $14^{th}$ symbols in this simulation setup. Linear interpolation is used to provide the CFR estimates in the remaining symbols.

A new realization is generated for each tower-to-UE channel in every frame. The CFO relative to each BS tower is randomly generated for every frame while staying within the limit of $10 \pm 0.2 $ ppm relative to the carrier frequency. As mentioned in Section III-\ref{S3a}, the tightest constraint on the BS oscillator accuracy is 0.1 ppm in the 5G specs. The current simulations exceed the error expected in such systems. Simulations are run for a minimum of 5000 frames or until twenty block errors are detected at a particular SNR. 

The CFOs corresponding to different towers  are estimated at the start of each frame with the help of a two-symbol preamble. The two symbols in this preamble are identical. Each preamble symbol consists of $N_T$ orthogonal bands in the frequency domain. Each band is a block of subcarriers with guard subcarriers in between. In the frequency domain, subcarriers of each band are loaded with a Zadoff-Chu sequence of appropriate length. 

The received time-domain samples corresponding to the two-symbol preamble are first filtered using a filter bank to separate the preamble signals belonging to each tower/band. The power of this filtered signal can be used to detect the presence or absence of an ITI signal, thus providing an estimate of $M$. This power can also be used to define the weights for performing WMD and as an indicator to select the appropriate estimator matrix $\mathbf{J}_M$. 

The timing and frequency synchronization procedure is implemented using the algorithm proposed in \cite{Schmidl1997}. The signal from the band corresponding to the desired tower can be cross-correlated with a time-domain preamble signal to estimate the timing offset of the signal from the desired tower. The first and second preamble symbols can be identified using this timing offset. The times of arrival of significant interferers are also estimated to determine an appropriate inter-block-interference (IBI)-free window. Since the two symbols will be identical except for the CFO and doppler induced phase shifts, the combined phase effect of CFO and doppler can be estimated by measuring the phase difference as suggested in \cite{Schmidl1997}. The proposed phase derotation scheme is now applied, followed by OFDM demodulation. 

The preamble sequence is not helpful for channel estimation since it is narrowband. Thus, the next step at the receiver is the channel estimation in the four symbols of a slot, followed by estimate interpolation over the entire slot. The channel estimation is done for the desired signal and the interfering signals. The estimates of the channel, noise variance, and CFOs are passed to the OC-JLLR detector to obtain the LLR values. The BLER is computed after LDPC decoding. 

The simulation results for various scenarios are presented in the following subsections.
\subsubsection{Low Mobility, $N_T=8$, $M=4$}
The desired signal is modulated with 4-QAM. Three ITI signals are also present with powers of 3, -3, and 0 dB relative to the desired signal, each modulated with 4-QAM. This means that the interferer powers are high and comparable to the desired signal power. The ITI signals are assumed to arrive at the time of flights given in Table \ref{Table2}. In pilot-carrying OFDM symbols, a comb-type pilot arrangement interlaced with data is assumed with a pilot subcarrier separation of 3. The number of used subcarriers and the number of pilots per OFDM symbol are given in Table \ref{Table2}. A TDL-B channel, as defined in \cite{3gpp38901} and scaled with an RMS delay spread of 300 ns (long delay spread channel), is assumed. In this scenario, the UE is assumed to have low mobility with speed restricted to below 10 kmph. 

\begin{figure}
\includegraphics[width=1\columnwidth]{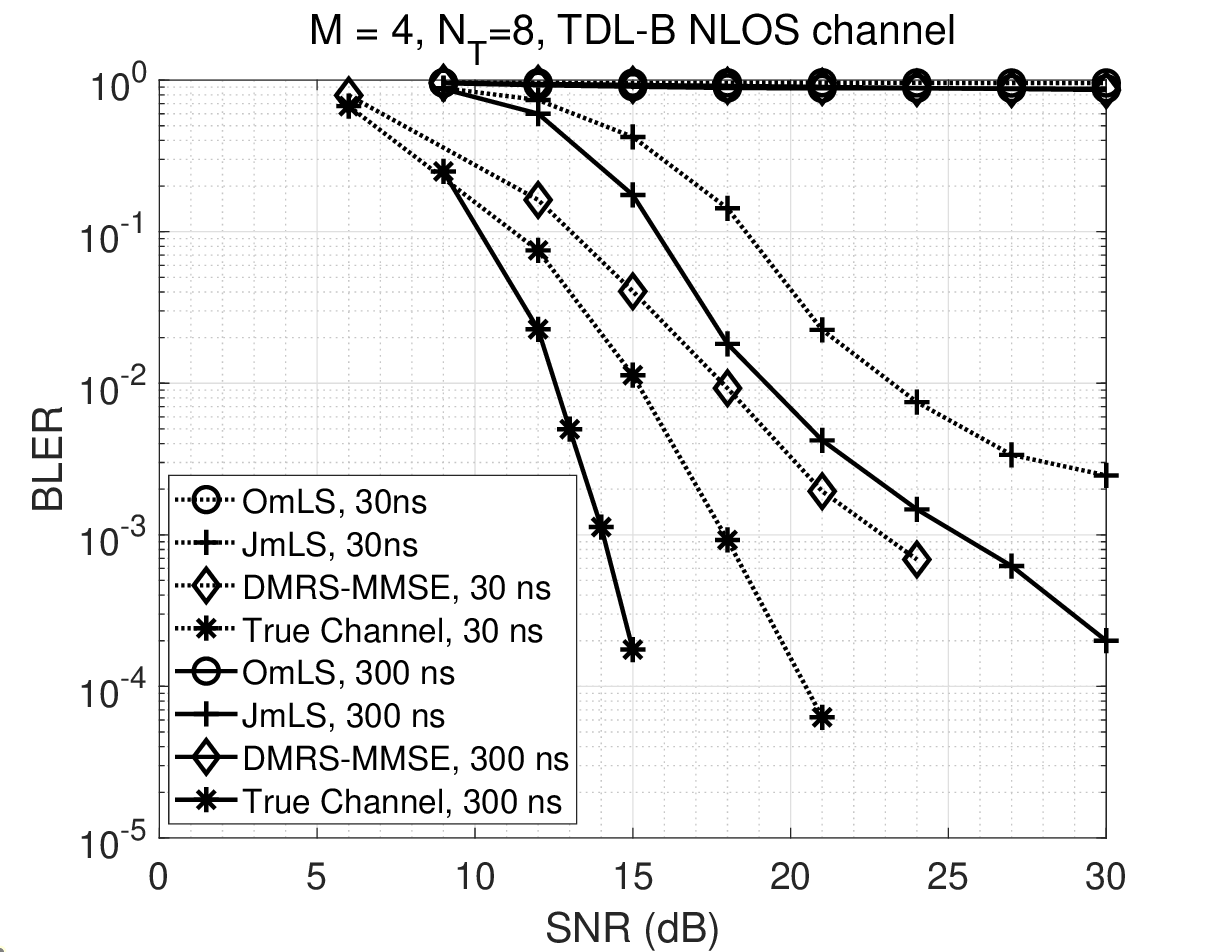}
\caption{BLER performance comparison of channel estimation schemes for $T_{rms} = 30$ ns and $T_{rms} = 300$ ns.}
\label{FigRC0}
\end{figure}

The performance of the OmLS, JmLS, and DMRS-MMSE channel estimation schemes are compared in terms of coded block error rate in Fig. \ref{FigRC0}. The SRS-DFT approach is not simulated here since its performance is found to be flooring similar to OmLS in Fig. \ref{FigRB2}. The performance of the joint detector fed with the true channel knowledge is also plotted for reference. 

It is seen that at low delay spreads, DMRS-MMSE is superior and is close to the performance for the case of true channel knowledge. This is unsurprising as the robust-MMSE interpolation is known to be near-optimal. The true values of the delay spread and SNR are given to the MMSE interpolator. Because of this, it can achieve near-ideal performance at low delay spreads. However, even at an rms delay spread of $300$ ns, DMRS-MMSE fails as the number of estimates becomes insufficient. 

By comparing Fig. \ref{FigRB2} and Fig. \ref{FigRC0}, it is seen that the BLER performance of JmLS improves at high delay spreads even though there is no improvement in the MSE performance. This is because high frequency selectivity is known to be beneficial for the performance of error correction codes \cite{Gacanin2008Frequency}. A similar BLER improvement is also seen in the case of true channel knowledge when the delay spread increases. 

The ICI compensation scheme proposed in this work comprises both a time-domain derotation scheme and the offset correction in the joint detector. It was found in Fig. \ref{FigICI1} that WMD performs close to the optimum time-domain derotator, and the MD performs close to the arctan derotator proposed in \cite{Deng2009Correction}. The SD has the worst performance among the time-domain derotators. The simulations results that follow shall only contain the results for SD, MD, and WMD schemes to avoid confusion. Further, only the OmLS and JmLS channel estimation schemes shall be included in future comparisons.

\begin{figure}
\includegraphics[width=1\columnwidth]{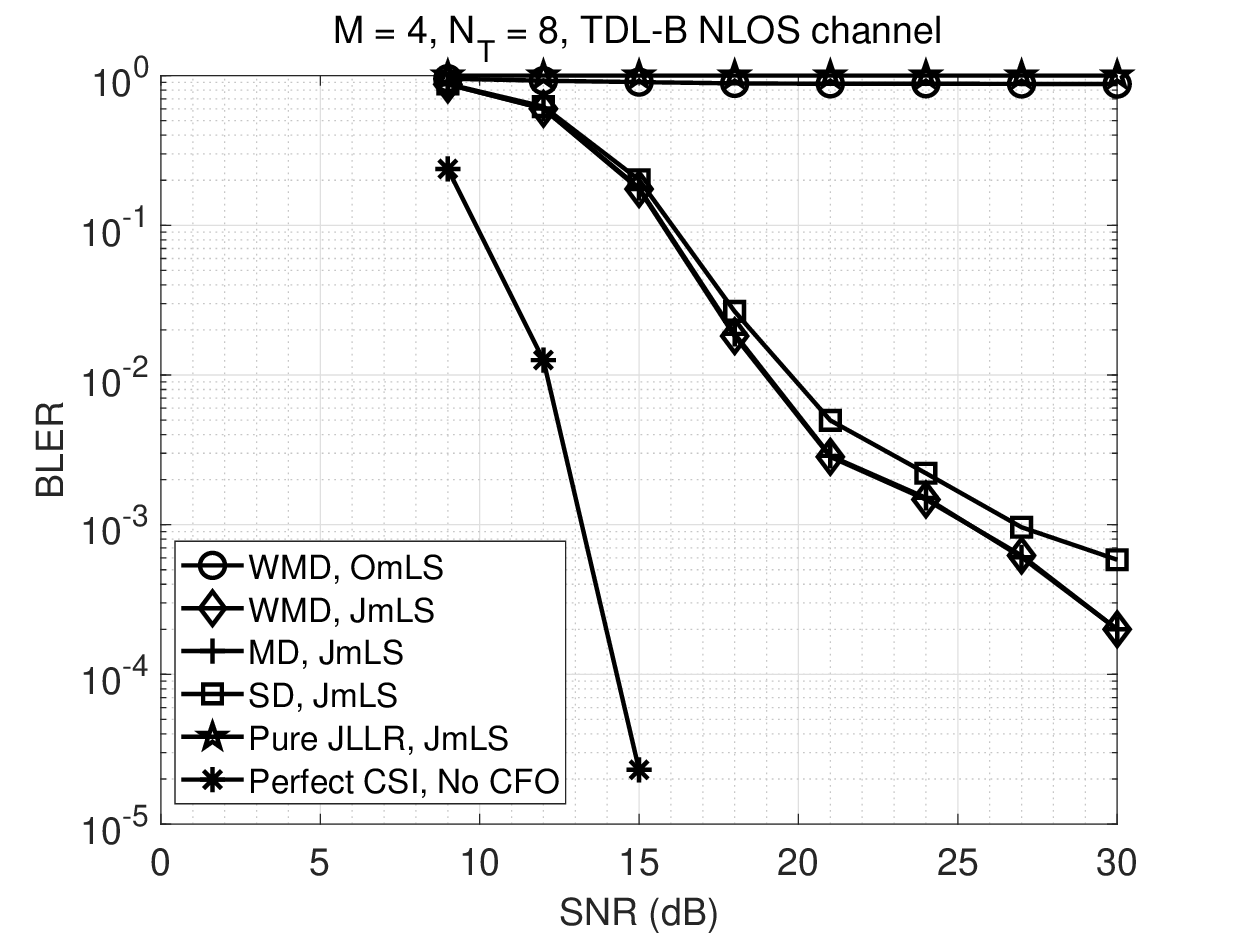}
\caption{Performance of joint reception for $M=4$ with ICI compensation}
\label{FigRC1}
\end{figure}

It is seen in Fig. \ref{FigRC1} that since the present design satisfies \eqref{JmLSSup}, the OmLS estimator has an unacceptable performance resulting in a BLER close to unity. This simulation shows that JmLS can perform well in scenarios where OmLS fails. The decoder fails in the absence of offset correction term as suggested by the proposed OC-JLLR in \eqref{eqOCJLLR}. All blocks will now be in error, irrespective of SNR. This is shown by the result indicated as ``pure JLLR" in Fig. \ref{FigRC1}. As a benchmark for comparison, the BLER when CFO is absent and when perfect channel knowledge is available at the receiver is also plotted.

The simulation results show that WMD and MD do not differ much in terms of BLER. The weights for WMD are almost equal in the cell-edge scenario, as the path losses from the different towers are comparable. It is seen that there is a difference of just 0.5 dB in terms of ICI cancellation for the two schemes. However, both perform better than the signal derotation (SD) approach used in interference-free systems. The SD approach floors more at high SNRs than WMD or MD. 

\subsubsection{Low Mobility, $N_T=8$, $M=3$, 16-QAM}
In this scenario, the desired signal is modulated with 16-QAM. Two ITI signals are also present with powers of 3 and -3 dB relative to the desired signal, each modulated with 4-QAM. The remaining assumptions in the previous section also hold here.
\begin{figure}
\includegraphics[width=1\columnwidth]{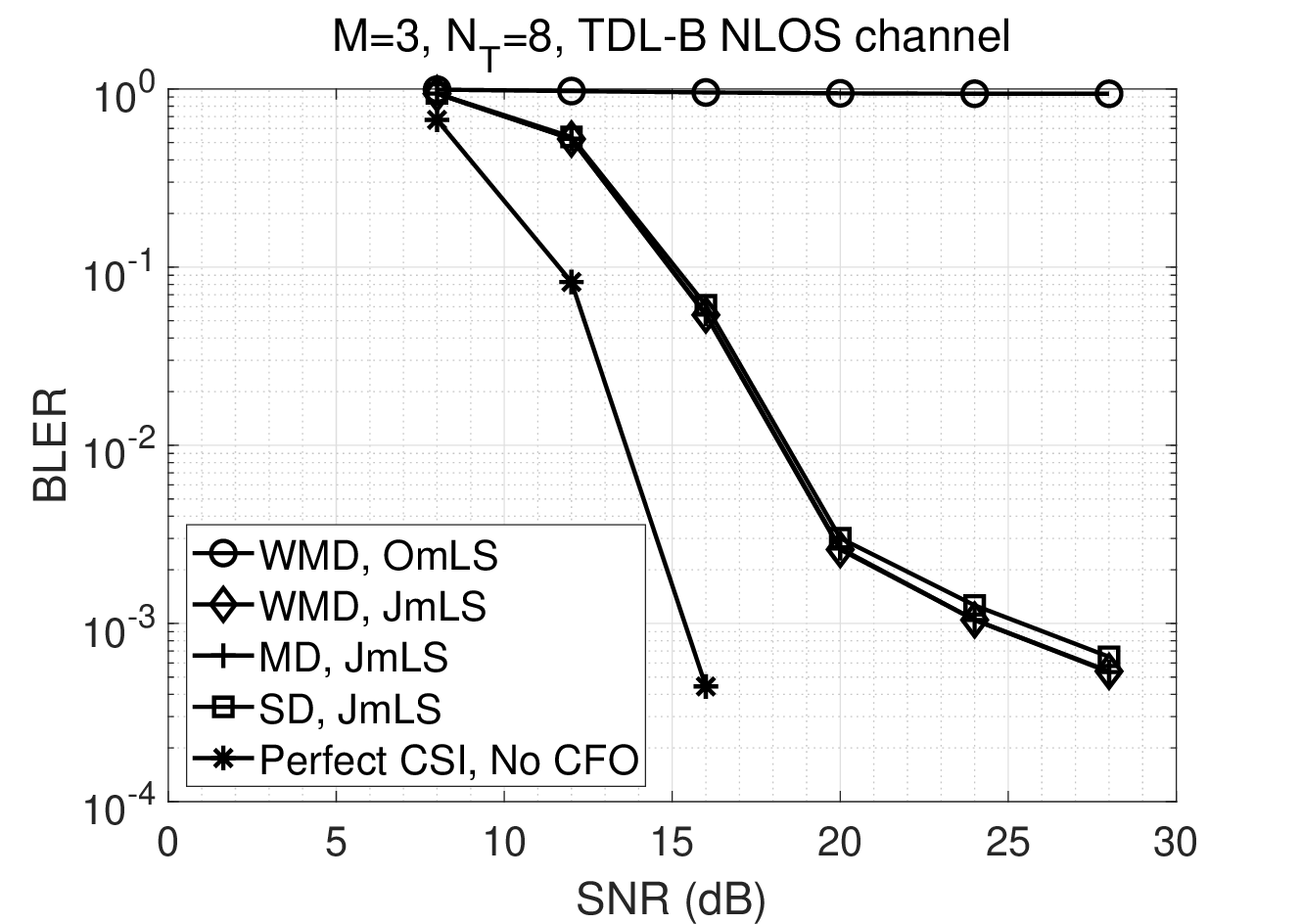}

\caption{BLER performance comparison for a 16-QAM desired signal with two 4-QAM interfering signals}
\label{FigRC2}
\end{figure}

The inferences drawn from this result (shown in Fig. \ref{FigRC2}) are similar to the $M=4$ case. It is seen that 16-QAM with two 4-QAM interferers perform slightly better than 4-QAM with three 4-QAM interferers even though the super-constellation order is the same. One reason for this behavior is that 16-QAM has better minimum-distance properties than the irregular super-constellation formed by adding two faded 4-QAM signals. In addition, the detector for $M=3$ requires only 3 sets of channel estimates and hence has less estimation error. 
\subsubsection{Low Mobility, $N_T=8$, $M=2$, 64-QAM}
\begin{figure}
\includegraphics[width=1\columnwidth]{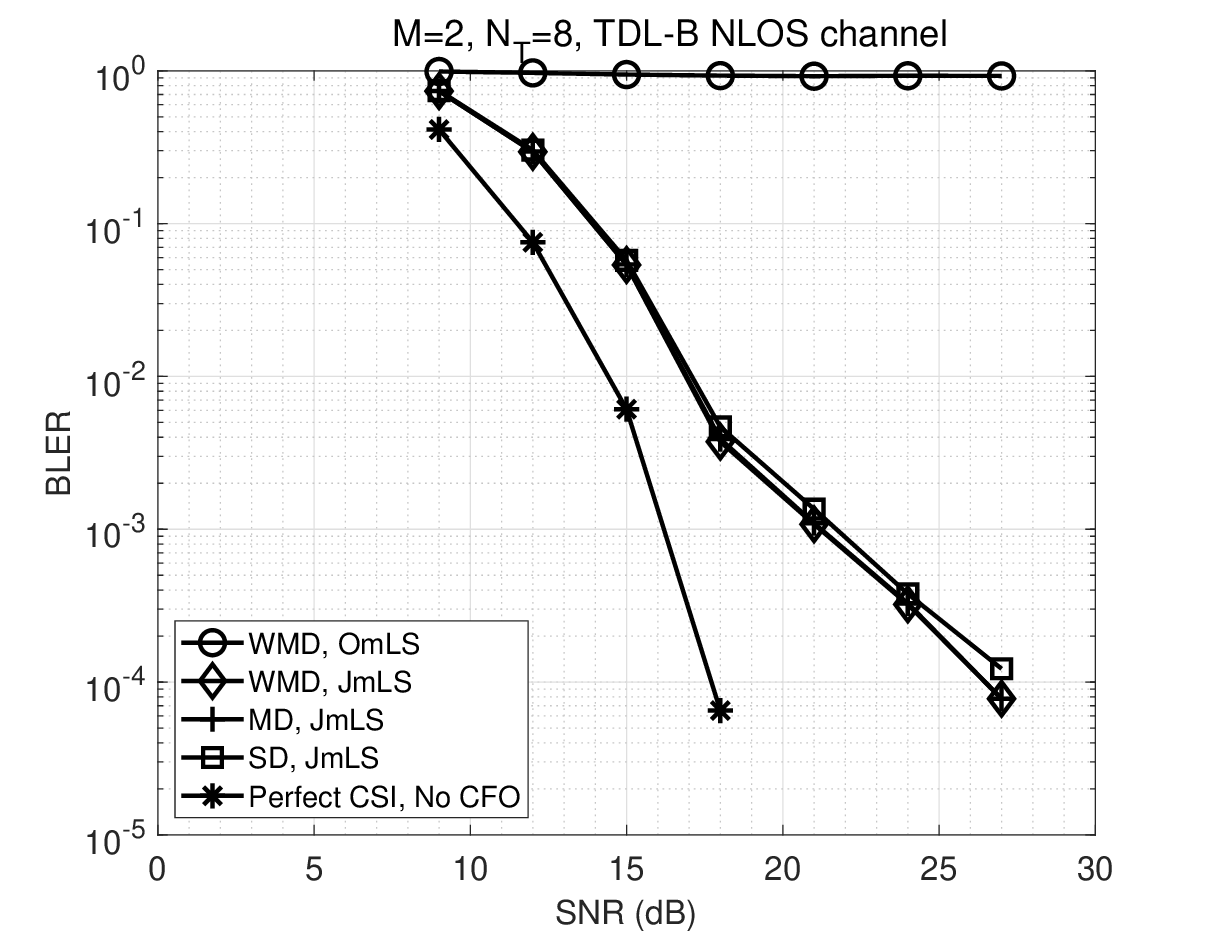}
\caption{BLER performance comparison for a 64-QAM desired signal with one 4-QAM interfering signal}
\label{FigRC3}
\end{figure}

The last scenario considered in the low mobility case is when the desired signal is modulated with 64-QAM. A single interfering signal whose power is 3 dB above the desired signal and modulated with 4-QAM is also present. This case is plotted in Fig. \ref{FigRC3}.

Although super-constellation remains the same, the performance for $M=2$ is superior to the previous two scenarios. This result also shows that a joint detector is feasible even for large constellation sizes like 64-QAM, but the number of interferers it can tolerate will reduce accordingly. For example, simulations show that it is not possible to accommodate two 4-QAM interferers on top of a 64-QAM desired signal for the code rate under consideration. Since high-order constellations are typically seen towards the center of the cell where interference is minimal, the results show that joint detectors can handle the interferences that are likely to be seen in the cell edge. 

\subsubsection{High Mobility, TDL-E LOS channel}
The final scenario uses a TDL-E channel model with a line-of-sight (LOS) Rician path and the usual non-line-of-sight (NLOS) Rayleigh multipaths. The UE is assumed to be mobile with a maximum speed of 100 kmph. The delay spread scaling factor considered for this channel is 100 ns since the TDL-E model is more spread out in terms of multipath. If a 300 ns scaling factor were to be considered as in the previous cases, this channel would have significant inter-block interference. 

\begin{figure}
\includegraphics[width=1\columnwidth]{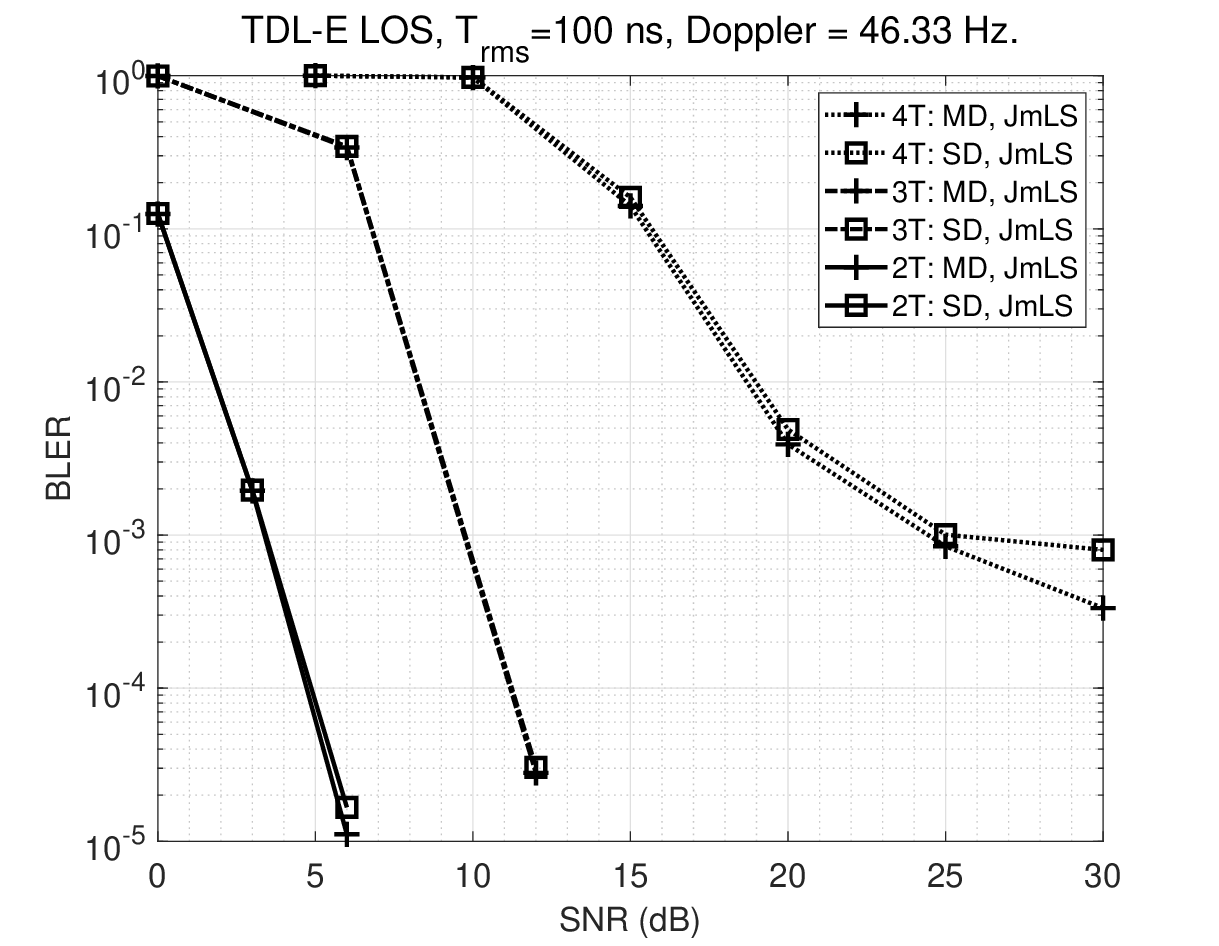}
\caption{BLER performance comparison in the high mobility case with a TDL-E LOS channel}
\label{FigRC4}
\end{figure}

The results plotted in Fig. \ref{FigRC4} consider a 4-QAM signal with 4-QAM interferer(s) for $M=$ 2, 3, and 4.  WMD is not plotted in these results since previous simulations have shown no significant difference between WMD and MD in terms of BLER for the given interference profile. The performance of OmLS is similar to the results shown so far and is not plotted here.
The $M=4$ result (denoted by 4T in the figure) shows that the proposed MD is superior to SD even in the presence of a significant Doppler component. Referring back to Fig. \ref{FigICI1}, it is seen that when the BS oscillator error is 0.2 ppm, the residual ICI power is around -30 dB for SD. This means that the ICI is comparable to noise at an SNR of 30 dB. Hence, SD starts to floor near 30 dB SNR. SD would floor sooner for higher CFOs, as predicted by Fig. \ref{FigICI1}. The effect of the derotation factor is not as pronounced for this particular code rate when $M<4$.

The results show that the SNR required to reach a particular BLER increases with increasing $M$. The first reason for this is an increase in the order of super-constellation. A secondary reason is the accumulation of errors in channel estimation of both the desired and interferer channels. This error accumulates linearly with $M$.

\section{Conclusion}
Interference-aware receiver algorithms that aid accurate signal recovery even in strong ITI were proposed in this work. It described a joint channel estimation scheme that uses a non-orthogonal pilot pattern to estimate the desired and interferer channels efficiently. The CRLB of the proposed estimator was derived and compared with its orthogonal counterpart. This joint estimator was shown to reduce the number of pilots by almost a factor of $N_T$. The condition under which the proposed estimator becomes superior to the orthogonal estimator was also derived, and the computational complexity of the estimator was analyzed. A low-complexity time-domain CFO compensation scheme that performs close to the optimal time-domain scheme was proposed. Although it does not have significant performance advantages over the existing arctan-based compensation, the proposed derotator can be  easily computed as it is a linear function of the CFOs. Irrespective of the time-domain derotator used, it should be coupled with a CFO correction factor to compensate for the phase shift caused due to the residual CFO. This compensation factor was derived for the JLLR detector in this work. 

Simulation results reveal that JmLS works with fewer pilots than OmLS, provided that condition \eqref{JmLSSup} is met. For the same pilot overhead, JmLS is shown to be advantageous at moderate to high delay spreads compared to an MMSE estimator that uses DM-RS pilots. All data blocks are erroneous in the absence of the proposed CFO correction factor for the JLLR. The proposed techniques enable high-quality signal recovery in interference-limited reuse-1 cellular OFDMA systems operating in the \mbox{sub-1 GHz} UHF band.
\begin{appendices}
\section{CRLB of joint channel estimate}
\label{Appendix:CRLBJmLS}
Ignoring the ICI term, the noise vector can be rewritten as
\begin{equation}
\mathbf{\tilde{W}} \approx \mathbf{Y}^{(D)}_p - \mathbf{X_p}\  \mathbf{F}_{pN_{cp}M} \ \mathbf{\tilde{h}'}_{N_{cp} M}
\end{equation}
Since $\mathbf{\tilde{W}}$ is a Gaussian vector, the log likelihood function of the data for the parameter $\mathbf{\tilde{h}'}_{N_{cp} M}$ is given by 
\medmuskip=0mu
\thinmuskip=0mu
\thickmuskip=0mu
\begin{align}
\begin{split}
&ln\ p(\mathbf{Y}^{(D)}_p;\mathbf{\tilde{h}'}_{N_{cp} M})\ =\ -\frac{N}{2} ln\ (2\pi\sigma^2)-\\ &\frac{(\mathbf{Y}^{(D)}_p - \mathbf{X_p} \mathbf{F}_{pN_{cp}M} \ \mathbf{\tilde{h}'}_{N_{cp} M})^H(\mathbf{Y}^{(D)}_p - \mathbf{X_p} \mathbf{F}_{pN_{cp}M} \ \mathbf{\tilde{h}'}_{N_{cp} M})}{2\sigma^2}
\end{split}
\end{align}
\medmuskip=3mu
\thinmuskip=4mu
\thickmuskip=5mu
The Fisher information matrix \cite{kay1993fundamentals} is then given by 
\begin{multline}
\mathbf{\mathcal{I}}(\mathbf{\tilde{h}'}_{N_{cp} M}) = -E \left[ \nabla_{\mathbf{\tilde{h}'}_{N_{cp} M}}^2 ln\ p(\mathbf{Y}^{(D)}_p;\mathbf{\tilde{h}'}_{N_{cp} M}) \right] \\= \frac{( {\mathbf{F}_{pN_{cp}M}}^H E\left[\mathbf{X_p}^H \mathbf{X_p}\right] \mathbf{F}_{pN_{cp}M})}{\sigma^2}.
\label{FisherInfoJmLS}
\end{multline}

CRLB of the total variance of an unbiased estimator of $\mathbf{\tilde{h}'}_{N_{cp} M}$ is found as the trace of the inverse of the Fisher information matrix \eqref{FisherInfoJmLS}.
\begin{equation}
CRLB_{tot}(\mathbf{\hat{h}_J}) = Tr\left( ({\mathbf{F}_{pN_{cp}M}}^H E\left[\mathbf{X_p}^H \mathbf{X_p}\right] \mathbf{F}_{pN_{cp}M})^{-1}  \right)\sigma^2
\label{ECRLBJmLStot}
\end{equation}
Here $\mathbf{\hat{h}_J}$ is an unbiased joint estimator of the CIR. Looking at \eqref{ECRLBJmLStot} for the case with unit-amplitude pilots such as 4-QAM, $\mathbf{X_i}^H\mathbf{X_i} = \mathbf{I}$. Although $\mathbf{X_i}^H\mathbf{X_j} \neq \mathbf{0}$ for any two integers $i \neq j,\ E\left[\mathbf{X_i}^H\mathbf{X_j}\right] = \mathbf{0}$. $\mathbf{X_i}^H\mathbf{X_j}$ is the product of two diagonal matrices with unit-amplitude sequences in the diagonal. This product results in another unit-amplitude sequence. The pilot sequences are assumed to be generated from a constellation with equal probability for all symbols. It then follows that the expected value of such a sequence is zero. Thus, $E[ \mathbf{X_p}^H \mathbf{X_p}] = \mathbf{I}$ and \eqref{ECRLBJmLStot} can be rewritten as
\begin{equation}
CRLB_{tot}(\mathbf{\hat{h}_J}) = Tr\left( ({\mathbf{F}_{pN_{cp}M}}^H \mathbf{F}_{pN_{cp}M})^{-1}  \right)\sigma^2
\label{ECRLBJmLSQAMpil}
\end{equation}
for the case of 4-QAM or any pilot sequence of unit amplitude. The above expression can be further simplified with the help of the following result. 
\begin{lemma}
For a subsampled DFT matrix $\mathbf{F}_{pN_{cp}}$ of size $N_p^{(J)} \times N_{cp}$, the product $\mathbf{F}_{pN_{cp}}^H  \mathbf{F}_{pN_{cp}} = N_p^{(J)} \mathbf{I}$, where $N_p^{(J)}$ is the number of joint estimator pilots.
\label{L1}
\end{lemma}
\begin{proof}
It follows from the definition of N-point DFT that $\mathbf{F}^H  \mathbf{F} = N \mathbf{I}$. Now, recall $\mathbf{F_{N_{cp}}}$ defined previously as a column subsampled version of $\mathbf{F}$. It is clear that any two columns of $\mathbf{F_{N_{cp}}}$ are orthogonal to each other. Thus, $\mathbf{F_{N_{cp}}}^H  \mathbf{F_{N_{cp}}} = N \mathbf{I_{N_{cp}}}$. Now, define  
\begin{align}
\begin{split}
\mathbf{F_{pN_{cp}}} = [F(i,k)], \quad & i = n p \text{ and } n = 0,1,2... \\ 
& k = 0,1... N_{cp}-1.
\end{split}
\end{align}
Here $p$ is the pilot spacing, i.e., number of subcarriers between consecutive pilots. 
\begin{equation}
F(i,k) = exp\left(\dfrac{-j 2 \pi k i}{N}\right) = exp\left(\dfrac{-j 2 \pi k n}{N/p}\right)
\label{DFTlemma}
\end{equation}
When the pilot spacing is such that the number of pilots, $N/p$, is an integer, \eqref{DFTlemma} shows that the N-point DFT becomes an $N/p$ point DFT. This happens when $p$ is a power of 2. In such cases, it follows that $\mathbf{F}_{pN_{cp}}^H  \mathbf{F}_{pN_{cp}} = \frac{N}{p} \mathbf{I}= N_p^{(J)} \mathbf{I}$.
\end{proof} 
This result is now applicable in the inverse term in \eqref{ECRLBJmLSQAMpil}.
\begin{align}
\left({\mathbf{F}_{pN_{cp}M}}^H \mathbf{F}_{pN_{cp}M}\right)^{-1} & = \frac{1}{N_p^{(J)}} \mathbf{I}_{MN_{cp} \times MN_{cp}}
\label{E16}
\end{align}

Then, \eqref{ECRLBJmLSQAMpil} can be rewritten as 
\begin{equation}
CRLB_{tot}(\mathbf{\hat{h}_J}) = \frac{M N_{cp}}{N_p^{(J)}} \sigma^2
\label{CRLBJmLS}
\end{equation}
This is a lower bound on the total variance across all channels.

\section{MSE of JmLS estimate}
\label{Appendix:MVUE}
The MSE of the joint estimate is given by the expression below.
\medmuskip=0mu
\thinmuskip=0mu
\thickmuskip=0mu
\begin{align}
\begin{split}
MSE_{_\Sigma}  &= Tr\left(E\left[(\mathbf{\hat{h}}_{JmLS} - \mathbf{\tilde{h}'}_{N_{cp} M}) (\mathbf{\hat{h}}_{JmLS} - \mathbf{\tilde{h}'}_{N_{cp} M})^H \right]\right)\\
 &= Tr\left(E\left[A+B+C+D\right]\right)
 \end{split}
\end{align}
\medmuskip=3mu
\thinmuskip=4mu
\thickmuskip=5mu
where,
\begin{align}
\begin{split}
A = &\mathbf{J}_M \left(\mathbf{X_p} \ \mathbf{F}_{pN_{cp} M} \ \mathbf{\tilde{h}'}_{N_{cp} M} +\mathbf{\mathbf{W}}\right) \times \\ &\left(\mathbf{X_p} \ \mathbf{F}_{pN_{cp} M} \ \mathbf{\tilde{h}'}_{N_{cp} M} +\mathbf{\mathbf{W}}\right)^H \mathbf{J}_M^H \\ = &\mathbf{J}_M \left(  \mathbf{X_p} \ \mathbf{F}_{pN_{cp} M} \mathbf{\tilde{h}'}_{N_{cp} M} {\mathbf{\tilde{h'}}_{N_{cp} M}}^H \mathbf{F}_{pN_{cp} M}^H \mathbf{X_p}^H+\right. \\ &\left. \hspace{-1.2cm}\mathbf{F}_{pN_{cp} M} \mathbf{W}^H + \mathbf{W} \mathbf{\tilde{h'}}_{N_{cp} M}^H  \mathbf{F}_{pN_{cp} M}^H \mathbf{X_p}^H +\mathbf{W}\mathbf{W}^H\right) \mathbf{J}_M^H 
\end{split}
\end{align}
The following equations simplify the expression for the expected value of A.
\begin{align}
\begin{split}
E[\mathbf{\tilde{h}'}_{N_{cp} M} \mathbf{\tilde{h'}}_{N_{cp} M}^H]  &= \sigma_h^2 \mathbf{I}\\
\mathbf{J}_M \mathbf{X_p} \mathbf{F}_{pN_{cp} M} &=\mathbf{F}_{pN_{cp} M}^H \mathbf{X_p}^H \mathbf{J}_M^H =   \mathbf{I}\\
E[\mathbf{W}] &= \mathbf{0}\\
E[\mathbf{W} \mathbf{W}^H]  &= \sigma^2 \mathbf{I}\\
\mathbf{J}_M \mathbf{J}_M^H &= \left( {\mathbf{F}_{pN_{cp} M}}^H \mathbf{X_p}^H \mathbf{X_p} \mathbf{F}_{pN_{cp} M}+\alpha \mathbf{I}\right)^{-1}
\end{split}
\label{ExpectedValues}
\end{align}
These relations can now be used on $A, B, C$ and $D$. 
\begin{align}
\begin{split}
&E[A]= \sigma_h^2 \mathbf{I}+\sigma^2 \left( {\mathbf{F}_{pN_{cp} M}}^H \mathbf{X_p}^H \mathbf{X_p} \mathbf{F}_{pN_{cp} M}+\alpha \mathbf{I}\right)^{-1}\\
&B = -\mathbf{J}_M \left(\mathbf{X_p} \ \mathbf{F}_{pN_{cp} M} \mathbf{\tilde{h}'}_{N_{cp} M} +\mathbf{\mathbf{W}}\right) \mathbf{\tilde{h'}}_{N_{cp} M}^H\\
&E[B]= -\sigma_h^2 \mathbf{I} \hspace{1.3cm} \left(\because \mathbf{J}_M \mathbf{X_p} \ \mathbf{F}_{pN_{cp} M} = \mathbf{I}\right)\\
&C = - \mathbf{\tilde{h}'}_{N_{cp} M} \left(\mathbf{X_p} \ \mathbf{F}_{pN_{cp} M} \mathbf{\tilde{h}'}_{N_{cp} M} +\mathbf{\mathbf{W}}\right)^H \mathbf{J}_M^H\\
&E[C]= -\sigma_h^2 \mathbf{I}\\
&D = \mathbf{\tilde{h}'}_{N_{cp} M} \mathbf{\tilde{h'}}_{N_{cp} M}^H\\
&E[D]= \sigma_h^2 \mathbf{I}
\end{split}
\end{align}
Thus,
\begin{align}
\begin{split}
MSE_{_\Sigma}  &= Tr\left(\sigma^2 \left( {\mathbf{F}_{pN_{cp} M}}^H \mathbf{X_p}^H \mathbf{X_p} \mathbf{F}_{pN_{cp} M}+\alpha \mathbf{I}\right)^{-1}\right)\\
&=\sigma^2 \frac{M N_{cp}}{N_p^{(J)}}
\end{split}
\end{align}
where the simplified expression (with $\mathbf{X_p}^H \mathbf{X_p} = \mathbf{I}$ ) is valid for unit-amplitude pilot sequences such as 4-QAM. As shown in Lemma 1 of Appendix \ref{Appendix:CRLBJmLS}, $\mathbf{F}_{pN_{cp} M}^H \mathbf{F}_{pN_{cp} M} = N_p^{(J)} \mathbf{I}$. As $\mathbf{I}$ is an $M N_{cp} \times M N_{cp}$ matrix here, its trace is equal to $M N_{cp}$.

\section{Time-domain derotation factor}
\label{Appendix:Derotation}
Referring to the system model described in \eqref{SII_E3} and the expression for ICI given in \cite{cho2010mimo}, the total ICI at the $k^{th}$ subcarrier is given by
\begin{align}
\begin{split}
ICI(k) &= \sum\limits_{m=0}^{M-1}C_m' e^{j2\pi\frac{N_{cp}}{N}\epsilon_m} \\ &\times \sum\limits_{l=0,l \neq k}^{N-1} \left(\frac{sin(\pi (l -k+ \epsilon_m) )}{Nsin(\pi \frac{(l -k+ \epsilon_m)}{N})}e^{j \pi (l-k+\epsilon_m)\frac{N-1}{N}}\right)
\\ &\times  H_{k,m} X_m[k] 
\end{split}
\label{Eqn:B1}
\end{align} 
The objective of reducing the ICI power has been realized in prior works \cite{Deng2009Correction,yang2015low} by employing a series of approximations and upper bounds on the ICI power. This paper uses a similar approach to obtain a different expression by employing different approximations. As a first step, it follows from the inequality of arithmetic and geometric means  \cite{chong1976arithmetic} that $a_1^2+a_2^2 \ge 2a_1a_2\ \forall\ a_1,a_2 \in 	\mathbb{C}$. This means that $(a_1+a_2)^2 \le 2(a_1^2+a_2^2)$. Repeated application of this property can be used to prove that $\left(\sum\limits_{m=0}^{M-1} a_m\right)^2 \le M \ \sum\limits_{m=0}^{M-1} a_m^2$. This upper bound can be applied to the expression for the ICI power at the $k^{th}$ subcarrier.
\begin{align}
|ICI(k)|^2 = &\left|\sum\limits_{m=0}^{M-1} f_1(\epsilon_m)\right|^2 \le M \sum\limits_{m=0}^{M-1} |f_1(\epsilon_m)|^2  \hspace{3cm}
\end{align}
where 
\medmuskip=0mu
\thinmuskip=0mu
\thickmuskip=0mu
\begin{align}
\begin{split}
|f_1(\epsilon_m)|^2 = &\left| \sum\limits_{l=0,l \neq k}^{N-1} \left(\frac{sin(\pi (l -k+ \epsilon_m) )}{Nsin(\frac{\pi (l -k+ \epsilon_m)}{N})}e^{j \pi (l-k+\epsilon_m)\frac{N-1}{N}}\right)\right|^2  \\ &\times |H_{k,m}|^2 |X_m[k]|^2 \text{\hspace{0.5cm}}
\end{split}
\end{align} 
\medmuskip=3mu
\thinmuskip=4mu
\thickmuskip=5mu
because $|ab|^2=|a|^2|b|^2$ and $|C_m' e^{j2\pi\frac{N_{cp}}{N}\epsilon_m}|=1$ as it is a pure phase term. A sinc function can be substituted for the ratio of sines in the above expression. This is possible because the FFT size $N$ is large and $sin (\frac{\theta}{N}) \approx \frac{\theta}{N}$. 

\begin{align}
\begin{split}
|f_1(\epsilon_m)|^2 = & \left| \sum\limits_{l=0,l \neq k}^{N-1} sinc(l -k+ \epsilon_m)e^{j \pi (l-k+\epsilon_m)\frac{N-1}{N}}\right|^2\\ & \times |H_{k,m}|^2 |X_m[k]|^2 \\ \le & \left( \sum\limits_{l=0,l \neq k}^{N-1} \left|sinc(l -k+ \epsilon_m)\right| \right)^2 |H_{k,m}|^2 |X_m[k]|^2
\end{split}
\end{align}
where the absolute value is taken inside the summation by employing the triangle inequality \cite{rudin1976principles}. It is now evident that the value of $\epsilon_m$ can be reduced to reduce the ICI. This is because, for $max(|\boldsymbol{\epsilon}|)\le 0.5$ and $l \neq k$, the absolute value of sinc reduces with decreasing $\epsilon_m$. On applying a time domain derotation factor $\epsilon_*$, the quantity $\epsilon_m$ in the equation gets replaced with $\epsilon_m-\epsilon_*$. We search for $\epsilon_*$ between $min(\boldsymbol{\epsilon})$ and $max(\boldsymbol{\epsilon})$ as the optimum value is known to lie within this range \cite{yang2015low}. Thus, the ICI power at the $k^{th}$ subcarrier is upper bounded by
\medmuskip=0mu
\thinmuskip=0mu
\thickmuskip=0mu
\begin{align}
\begin{split}
|ICI(k)|^2 \le  & \sum\limits_{m=0}^{M-1} \left( \sum\limits_{l=0,l \neq k}^{N-1} \left|sinc(l -k+ \epsilon_m-\epsilon_*)\right| \right)^2 f_2 (k,m) \\ f_2 (k,m) =& M |H_{k,m}|^2 |X_m[k]|^2
\end{split}
\end{align}
\medmuskip=3mu
\thinmuskip=4mu
\thickmuskip=5mu
The inequality of arithmetic and geometric means \cite{chong1976arithmetic} can once again be used to upper bound the above square of a summation with a scaled version of summation of squares.
\medmuskip=0mu
\thinmuskip=0mu
\thickmuskip=0mu
\begin{align}
\begin{split}
|ICI(k)|^2 \le  & \sum\limits_{m=0}^{M-1} \ \sum\limits_{l=0,l \neq k}^{N-1} \left(sinc(l -k+ \epsilon_m-\epsilon_*)\right)^2 f_3 (k,m) \\ f_3 (k,m) =& M N |H_{k,m}|^2 |X_m[k]|^2
\end{split}
\end{align}
\medmuskip=3mu
\thinmuskip=4mu
\thickmuskip=5mu
By following the approximation used in \cite{Deng2009Correction} and \cite{Pollet1995BER} for large $N$, $\sum\limits_{l=0,l \neq k}^{N-1} \left(sinc(l -k+ \epsilon_m-\epsilon_*)\right)^2 \approx 1-$ \hbox{$sinc^2(\epsilon_m-\epsilon_*))$}. Since $N$ is the total number of subcarriers in the OFDM system, it is usually of the order of $10^3$ and the approximation will hold. The next step is to find the optimum derotator $\epsilon_*$ that minimizes the upper bound on ICI power given above. Then, the cost function is
\begin{align}
C(\epsilon_*) =  & \sum\limits_{m=0}^{M-1} (1- sinc^2(\epsilon_m-\epsilon_*))f_3 (k,m)
\label{CostFn}
\end{align}
Using the first two terms of the Maclaurin expansion for $sinc(x)$, the approximation $sinc(x) \approx 1-\frac{(\pi x)^2}{3!}$ can be made. Then $sinc^2(x) \approx 1-2\frac{(\pi x)^2}{3!}+(\frac{(\pi x)^2}{3!})^2$. Here, $x=\epsilon_m-\epsilon_*$. The higher order terms of the Maclaurin expansion will become negligible when the sinc function is squared. Substituting this expansion in \eqref{CostFn},
\medmuskip=0mu
\thinmuskip=0mu
\thickmuskip=0mu
\begin{align}
C(\epsilon_*) =  & \sum\limits_{m=0}^{M-1} \left(2\frac{(\pi (\epsilon_m-\epsilon_*))^2}{3!}-\frac{(\pi^4 (\epsilon_m-\epsilon_*)^4)}{(3!)^2}\right)f_3 (k,m)
\end{align}
Differentiating the cost function, 

\begin{align}
\begin{split}
\frac{\partial C(\epsilon_*)}{\partial \epsilon_*}  &=   \sum\limits_{m=0}^{M-1} \left(\frac{-4\pi (\epsilon_m-\epsilon_*)}{3!}+\frac{4(\pi^4 (\epsilon_m-\epsilon_*)^3)}{(3!)^2}\right)f_3 (k,m) \\
&=\sum\limits_{m=0}^{M-1} \left(-((\epsilon_m-\epsilon_*))+\frac{(\pi^3 (\epsilon_m-\epsilon_*)^3)}{3!}\right)f_3 (k,m)
\end{split}
\label{diffCostFn}
\end{align}
\medmuskip=3mu
\thinmuskip=4mu
\thickmuskip=5mu

In a 5GNR-type deployment, the oscillator at the UE has significantly lower accuracy requirements than the BS \cite{Lin2018Synchronization}. Oscillator accuracy at the UE is of the order of 10 parts per million (ppm) of the carrier frequency \cite{Yuichi2020Epson} and as low as 0.1 ppm at the base station side \cite{Lin2018Synchronization}. Thus, the CFO will have the same sign for all the ITI terms. The CFO of the different towers will vary between 9 ppm and 11 ppm if a BS CFO of 1 ppm is considered. For the maximum carrier frequency of 1 GHz (as sub-1 GHz systems are considered in this work) and a subcarrier spacing of 15 kHz, the normalized CFO would be between $0.6$ and $0.73$. This means that $\epsilon_m-\epsilon_*$ is of the order of 0.1. The cubed term in \eqref{diffCostFn} will then be two orders of magnitude lower than the linear term and can hence be ignored while finding the minima. Substituting for $f_3 (k,m)$ and equating the derivative to zero, 
\begin{align}
&\sum\limits_{m=0}^{M-1} \left(\epsilon_m-\epsilon_*\right) |H_{k,m}|^2 |X_m[k]|^2=0
\end{align}
Since the derotation is done in the time domain, the dependence on the index $k$ needs to be removed. Thus, the terms dependent on $k$ are averaged over all subcarriers.
\begin{align}
\begin{split}
& \sum\limits_{m=0}^{M-1} \left(\epsilon_m-\epsilon_*\right) \frac{1}{N}\sum\limits_{k=0}^{N-1}  |H_{k,m}|^2 |X_m[k]|^2=0\\
\implies & \sum\limits_{m=0}^{M-1} \left(\epsilon_m-\epsilon_*\right) P_m = 0
\end{split}
\end{align}
where $P_m$ is the power of the $k^{th}$ channel scaled by the mean value of the constellation of the $m^{th}$ tower. This mean value is independent of the index $m$ as it will be common across all the towers. This is because the transmitted power per OFDM symbol is the same for all base stations. Thus, 
\begin{align}
\epsilon_{*,opt} = \frac{\sum\limits_{m=0}^{M-1} \epsilon_m P_m}{\sum\limits_{m=0}^{M-1} P_m}
\end{align}

\section{Offset Corrected Joint Detector}
\label{Appendix:JLLR}
Bayes Theorem is applied to the joint detector formulation described in \eqref{LLRproblem}. 
\medmuskip=0mu
\thinmuskip=0mu
\thickmuskip=0mu
\begin{align}
\begin{split}
&LLR_{0,\lambda,k} = \hfill \\
&ln \left( \dfrac{P\left(Y_D[k]\; \middle| \;b_\lambda \left( X_0[k]\right) = 1,\mathbf{\tilde{C}}', \mathbf{\tilde{H}}_k'\right)P(b_\lambda \left( X_0[k]\right) = 1)}{P\left(Y_D[k]\; \middle| \;b_\lambda \left( X_0[k]\right) = 0,\mathbf{\tilde{C}}', \mathbf{\tilde{H}}_k'\right)P(b_\lambda \left( X_0[k]\right) = 0)} \right)
\end{split}
\end{align}

 Summation is done over all possibilities for which $b_\lambda \left( X_0[k]\right) = 1$ in the numerator and $b_\lambda \left( X_0[k]\right) = 0$ in the denominator,assuming equal prior probabilities for constellation points. That is, the marginal distribution is found from the joint distribution.  
\begin{align}
\begin{split}
&LLR_{0,\lambda,k} = \\
&ln \left( \dfrac{ \sum\limits_{X_0 \in \boldsymbol{X_0^{(1,\lambda)}}}  \sum\limits_{ \substack{X_m \in \boldsymbol{X_m}, \\m \neq 0}} P\left(Y_D[k]\; \middle| \;\boldsymbol{C}, \boldsymbol{\tilde{H}}_k', \boldsymbol{X[k]}\right)}{\sum\limits_{X_0' \in \boldsymbol{X_0^{(0,\lambda)}}}  \sum\limits_{ \substack{X_m \in \boldsymbol{X_m}\\ m \neq 0}} P\left(Y_D[k]\; \middle| \;\boldsymbol{C}, \boldsymbol{\tilde{H}}_k', \boldsymbol{X[k]}\right)} \right)
\end{split}
\end{align}
$\boldsymbol{X_0^{(1,\lambda)}}$ denotes the subset of the constellation of $\boldsymbol{X_0}$ for which the $\lambda^{th}$ bit is 1, and $\boldsymbol{X_0^{(0,\lambda)}}$ denotes the subset of the constellation of $\boldsymbol{X_0}$ for which the $\lambda^{th}$ bit is 0.
Assuming that the residual ICI does not distort the Gaussianity of the noise heavily, the JLLR equation becomes:

\begin{align}
\begin{split}
&LLR_{0,\lambda,k} = \\ &ln \left\lbrace \dfrac{\sum\limits_{\substack{X_0 \in \boldsymbol{\boldsymbol{X_0^{(1,\lambda)}}}\\X_m \in \boldsymbol{X_m}\\ m \neq 0}} exp \left(  \begin{array}{l} -\frac{1}{\sigma^2}\|
Y_D[k] - \tilde{C}'_0 \tilde{H}_{k,0}X_0[k]  + \\  \ \ \quad \sum\limits_{\substack{m \in \mathcal{M}\\ m \neq 0}} \tilde{C}'_m \tilde{H}_{k,m}X_m[k] \|^2   \end{array} \right) }{\sum\limits_{\substack{X_0' \in \boldsymbol{\boldsymbol{X_0^{(0,\lambda)}}}\\X_m \in \boldsymbol{X_m}\\ m \neq 0}} exp \left( \begin{array}{l} -\frac{1}{\sigma^2}\|  Y_D[k] - \tilde{C}'_0 \tilde{H}_{k,0}X_0'[k]  + \\ \ \ \quad \sum\limits_{\substack{m \in \mathcal{M}\\ m \neq 0}} \tilde{C}'_m \tilde{H}_{k,m}X_m[k] \|^2 \end{array} \right)} \right\rbrace
\end{split}
\end{align}
The complexity of the detector can be reduced by using the suboptimal Max-Log-MAP approximation proposed in \cite{robertson1997optimal},\\
\medmuskip=0mu
\thinmuskip=0mu
\thickmuskip=0mu
\begin{multline}
LLR_{0,\lambda,k} \approx  \min\limits_{\substack{X_0 \in \mathbf{X}_0^{\lambda_1}, \\X_m \in \mathbf{X_m}\\m \neq 0}}  \frac{1}{\sigma^2}\| Y_D[k] - \tilde{C}'_0 \ \tilde{H}_{k,0}X_0[k] \\ - \sum\limits_{m=1}^{M-1} \tilde{C}'_m \  \tilde{H}_{k,m}X_m[k] \|^2 \\ 
 - \min\limits_{\substack{X_0' \in \mathbf{X}_0^{\lambda_0},\\X_m[k] \in \mathbf{X_m}\\m \neq 0} } \frac{1}{\sigma^2}\|  Y_D[k] - \tilde{C}'_0 \ \tilde{H}_{k,0}X_0'[k]  \\- \sum\limits_{m=1}^{M-1} \tilde{C}'_m \  \tilde{H}_{k,m} X_m[k] \|^2
 \label{eqOCJLLR_app}
\end{multline}
This expression shall be called the offset-corrected joint log likelihood ratio (OC-JLLR) expression. 

\end{appendices}

%

\section*{ACKNOWLEDGMENT}
The authors would like to thank the anonymous peer reviewers for their valuable feedback.
\bibliographystyle{IEEEtran}
\bibliography{IEEEabrv,paperrefs_new}

\begin{IEEEbiography}[{\includegraphics[width=1in,height=1.25in,clip,keepaspectratio]{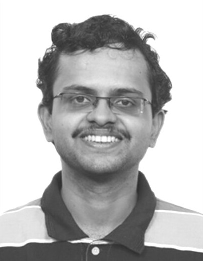}}]{ABHAY MOHAN M V }   received the B.Tech degree in electronics and communication engineering from the Government Engineering College-Palakkad, Kerala, India, in 2014 and the M.Tech degree in Communication Engineering and Signal Processing from the Government Engineering College-Thrissur, Kerala, India, in 2017. He is currently pursuing
the Ph.D. degree in electrical engineering from the Indian Institute of Technology Madras, Chennai, India. 
\end{IEEEbiography}

\begin{IEEEbiography}[{\includegraphics[width=1in,height=1.25in,clip,keepaspectratio]{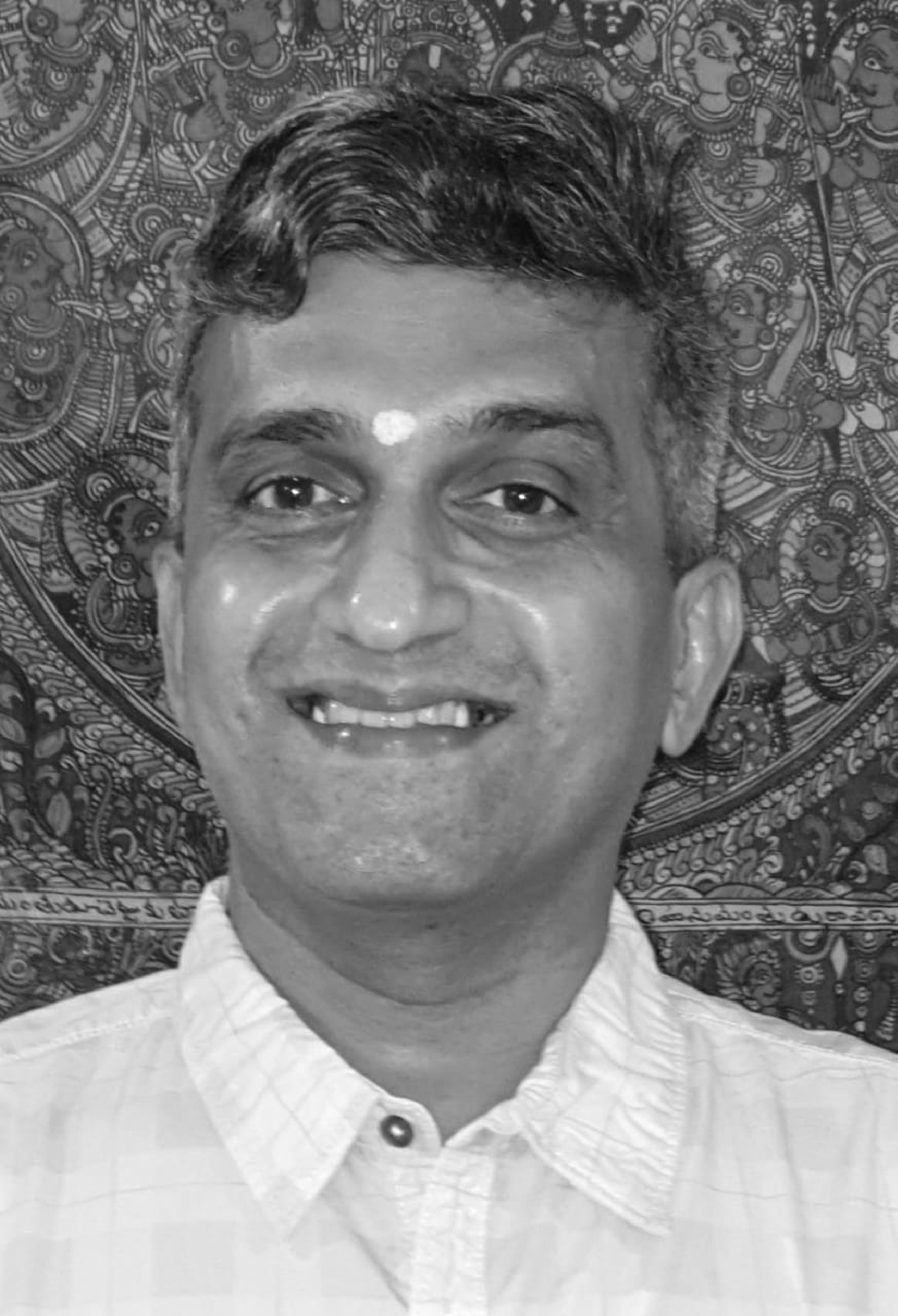}}]{K. GIRIDHAR } received the B.Sc. degree in applied sciences from the PSG College of Technology, Coimbatore, India, in 1985, the ME degree in electrical communications from the Indian Institute of Science, Bangalore, India, in 1989 and the PhD degree in electrical engineering from the University of California, Santa Barbara, USA in 1993. 

He was a member of research staff at CRL, Bharat Electronics, Bangalore, India, between 1989-90 and a research affiliate in electrical engineering at Stanford University, California, USA, between 1993-94. He is currently a Professor at the Indian Institute of Technology Madras (IITM), Chennai, India. His research interests are broadly in the areas of adaptive signal processing and wireless communications systems, with an emphasis on various transceiver algorithms and custom air-interface design for strategic communications. More recently, his research group has focused on non-orthogonal spectrum sharing and distributed radar signal processing.

Dr. Giridhar has been a consultant for many telecom and signal processing industries and research labs in India, and has taken sabbaticals with Gigabit Communications Inc, San Jose in summer of 2000, with Beceem Communications Inc, Santa Clara, between 2004-05, and with Data Patterns (India) Ltd, Chennai between 2018-19. He is a founding Director of QallCover Communications LLP, Chennai, and has been a visiting faculty at Sri Sathya Sai Institute of Higher Learning, Prasanthi Nilayam, and at Stanford University.
\end{IEEEbiography}

\end{document}